\documentclass[twocolumn,prb]{revtex4-2} 
\usepackage{amsfonts}
\usepackage[T1]{fontenc}
\usepackage{amsmath,amsbsy,amssymb,graphicx}
\usepackage{times}

\def\red#1{\textcolor{red}{#1}}

\let\red=\relax
\begin{document}

\title{Ising Machine based on Bistable Microelectromechanical Systems}
\author{Motohiko Ezawa$^1$, Eric Lebrasseur$^2$ and Yoshio Mita$^2$}
\affiliation{$^1$Department of Applied Physics, University of Tokyo, Hongo 7-3-1,
113-8656, Japan}
\affiliation{$^2$Department of Electrical Engineering, University of Tokyo, Hongo 7-3-1,
113-8656, Japan}

\begin{abstract}
We propose an Ising machine made of microelectromechanical systems (MEMS),
where the annealing process is automatically executed by a dissipation
mechanism. The core structure is a series of buckled plates. Two stable
positions of each plate (left and right) represent its binary state acting
as a bit so that a plate works as a mechanical memory. The electrostatic
interaction between adjacent plates is introduced by applying voltage.
Plates continue to flip between two stable buckled positions until the
series of plates reaches a local minimum due to the damping of the
mechanical motion. \red{First, we design Ising machines simulating a
ferromagnetic (FM) interaction and an antiferromagnetic (AF) interaction
separately. Then, we propose a fully-connected MEMS\ network representing a
coexistence system of FM and AF interactions in an arbitrary way, by way of
which an arbitrary combinatorial problem described by the Ising model can be
solved.} The present mechanism works at room temperature without external
magnetic field, which is very different from the standard classical or
quantum annealing mechanism.
\end{abstract}

\maketitle

\section{Introduction}

Many binary-combinatorial problems are described by the Ising model\cite%
{Lucas}. It is solved by an Ising machine, where a local minimum state is
efficiently obtained by an annealing method. There are various ways for the
annealing process. A classical way is a simulated annealing\cite{Kirk,Cerny}%
, where we decrease temperature to obtain a local minimum state. The quantum
annealing machine\cite{Nishimori,Farhi,Johnson} is one of the most
successful quantum machines on a commercial basis although it is not a
universal quantum computer. It determines a local minimum state of a
transverse-field Ising model by gradually decreasing the magnetic field.
However, it is necessary to cool down the sample below the helium
temperature in order to make the system superconducting.

Recently, many variations of the Ising machine are proposed\cite{Afoa} such
as a coherent Ising machine\cite{Yamamoto,Wang,Mara,Inagaki,Bohm} based on
photonic systems, an electronic oscillator-based Ising machine\cite{OIM},
accelerated simulated annealers\cite{Yamaoka,Takemoto}, a simulated
bifurcation machine\cite{Goto,Goto2} based on digital computer and a
probabilistic-bit (p-bit) machine\cite{Ohno,Cam,Cam2,Per,Cam3} based on
magnetic systems. The key point is how to make a binary stable state
representing the Ising degree of freedom and how to control it. It is
advantageous if we are able to construct an Ising machine with the use of
current electromechanical technologies.

There is an important observation made by Feynman\cite{Feynman}. He proposed
to use the equations of motion describing a natural phenomenon to solve a
given problem in the context of quantum computations. The execution is done
all at once at high speed. It is contrasted to the digital computer, where
the calculation is done sequentially.

A microelectromechanical system (MEMS) is a combinatorial system of electric
circuits and mechanical systems\cite{Mita1,Mita2,Toshiyoshi}. A typical
example is an electrostatic actuator consisting of parallel plates
abbreviated as a parallel-plate MEMS. Another important MEMS\ is a
comb-teeth electrostatic actuator abbreviated as a comb-teeth MEMS.

In this paper, we propose an Ising machine with the use of a buckled-plate
MEMS made of many parallel plates, which is a kind of a parallel-plate MEMS.
\red{We start with Ising machines simulating a FM interaction and an AF
interaction separately. Then, we present a fully-connected MEMS\ network
containing both FM and AF interactions in an arbitrary way. It enables us to
solve an arbitrary combinatorial problem described by the Edwards-Anderson
model\cite{Edwards}, where the FM and AF interactions coexist randomly.}
The annealing process is shown to be executed automatically due to the
damping of the plate motion based on the analytical study of the
Euler-Lagrange-Rayleigh equation, where all plates stop once the system
reaches a local minimum. We also demonstrate numerically how the annealing
actually proceeds. We emphasize that the present Ising machine works at the
room temperature without external magnetic field. It is highly contrasted to
simulated annealing and quantum annealing, where it is necessary to
precisely control external parameters such as temperature and magnetic field.

This paper is composed as follows. In Sec.II we study a buckled-plate MEMS
made of many parallel plates, where each plate has two stable positions. In
Sec.III, we realize an Ising model by this buckled-plate MEMS with
sequential parallel plates to form a chain. We show that the ground state is
FM. Buckled plates are manipulated by controlling voltage. We analytically
prove that the system reaches a local minimum based on the
Euler-Lagrange-Rayleigh formalism. In Sec.IV, we propose an AF Ising model
by introducing seesaw structures. \red{In Sec.V, we propose an Ising machine
materialized by a fully-connected MEMS network simulating both FM and AF
interactions.} Sec. VI is devoted to discussion. We prepare Appendices,
where we derive various equations used in the main text. We also discuss an
Ising model with the use of comb-teeth MEMS by controlling charge.

\begin{figure}[t]
\centerline{\includegraphics[width=0.48\textwidth]{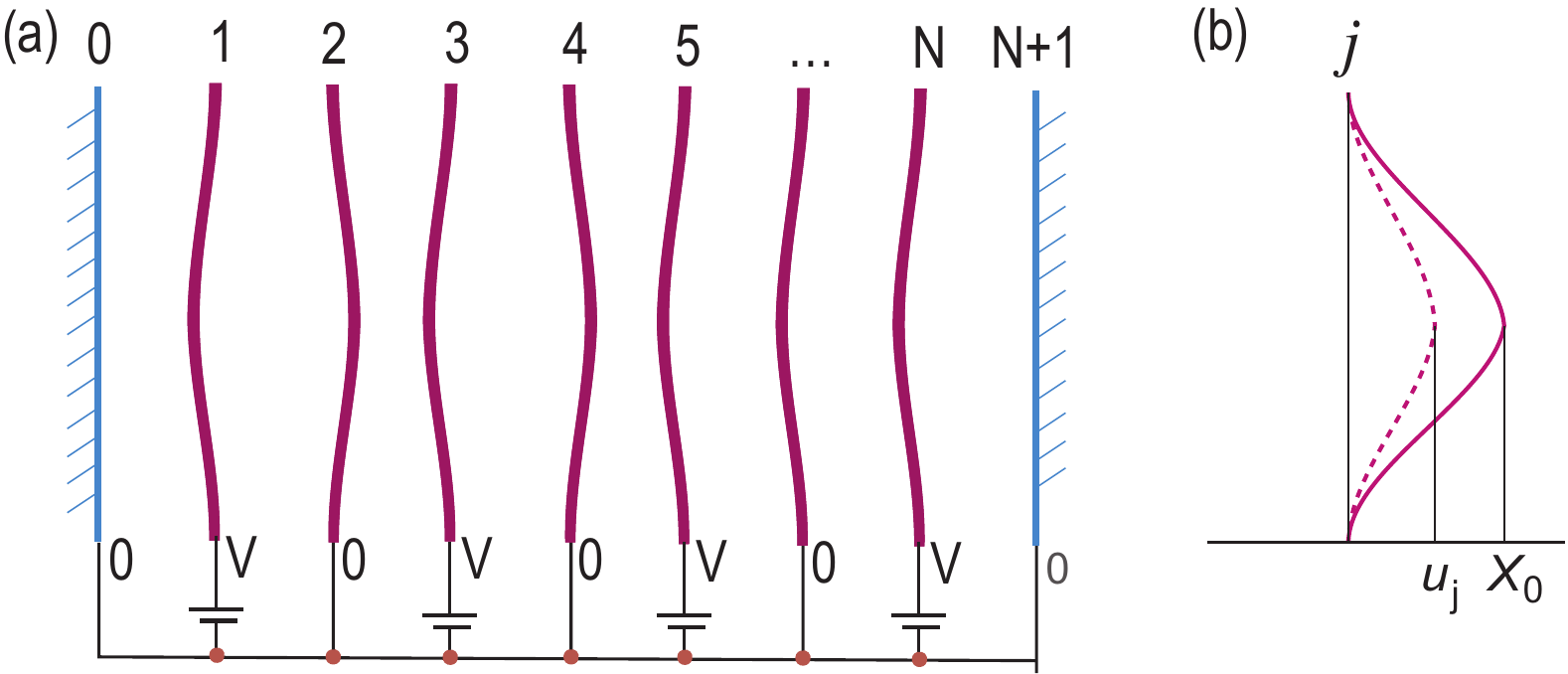}}
\caption{ (a) Illustration of a buckled-plate MEMS, where the direction of
buckling is controlled by the electrostatic interaction between the adjacent
plates. The voltage at each plate is either $0$ or $V$, alternatively . (b)
Enlarged illustration of the $j$-th buckled plate with the plate position $%
u_j$ and the stable buckled position $X_0$ indicated. }
\label{FigBoundary}
\end{figure}

\begin{figure*}[t]
\centerline{\includegraphics[width=0.98\textwidth]{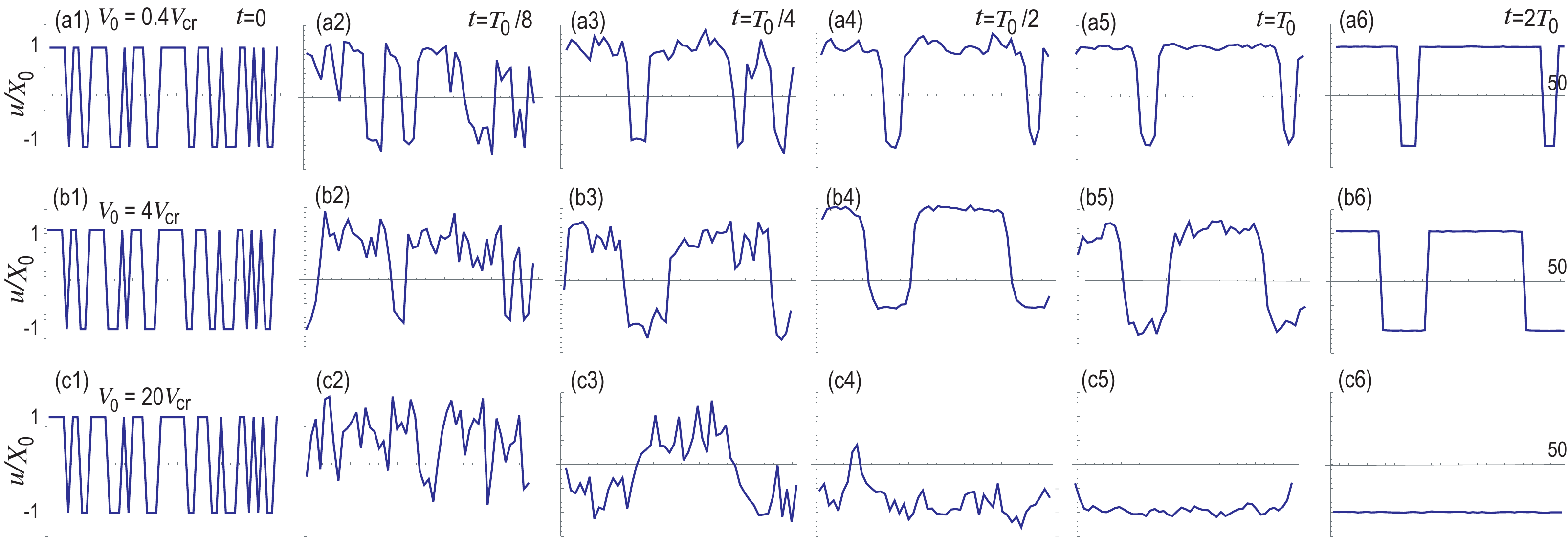}}
\caption{ Dynamics with the fixed-boundary condition. (a1)$\sim $(c1)
Configuration $u_{j}/X_{0}$ at the initial state $t=0$; (a2)$\sim $(c2) that
at $t=T_{0}/8$; (a3)$\sim $(c3) that at $t=T_{0}/4$; (a4)$\sim $(c4) that at 
$t=T_{0}/2$; (a5)$\sim $(c5) that at $t=T_{0}$; (a6)$\sim $(c6) that at $%
t=2T_{0}$. We have used a 50-plate system. (a1)$\sim $(a6) The applied
voltage is $V=0.4V_{\text{cr}}$; (b1)$\sim $(b6) it is $V=4V_{\text{cr}}$;
(c1)$\sim $(c6) it is $V=20V_{\text{cr}}$. (a1)$\sim $(a6) The applied
voltage is $V=0.4V_{\text{cr}}$; (b1)$\sim $(b6) it is $V=4V_{\text{cr}}$;
(c1)$\sim $(c6) it is $V=20V_{\text{cr}}$. (a1)$\sim $(a6) There is no
annealing when the voltage is below the critical value ($V<V_{\text{cr}}$).
(b1)$\sim $(b6) There remain two domain walls after annealing when the
voltage exceeds the critical value but not sufficiently. (c1)$\sim $(c6) The
annealing process is perfect when the voltage exceeds the critical value
sufficiently. The vertical axis is the configuration $u_{j}/X_{0}$. The
horizontal axis is the position $j$. We have set $\protect\alpha =0.02$, $%
\protect\gamma =0.01$, $T_{0}=500$, $X_{0}=1$, $m=1$ and $\protect\varepsilon%
_0 S=1$. }
\label{FigDynamicsF}
\end{figure*}

\section{Buckled-Plate MEMS}

In the field of MEMS, the bistable structure has been studied with a typical
application to memories\cite{Vang, Inta}. We consider a system made of $N$
plates placed parallelly with equal separation, where all plates are buckled
as in Fig.\ref{FigBoundary}(a). A buckled plate carries the classical bit
information $+1$ ($-1$), when it is buckled rightward (leftward). All plates
are buckled either rightward or leftward in the FM ground state.

The potential energy of the system is well approximated by\cite{Rincon} 
\begin{equation}
U_{\text{mech}}=\sum_{j=1}^{N}\frac{\alpha }{2}\left(
u_{j}^{2}-X_{0}^{2}\right) ^{2},  \label{Pote}
\end{equation}%
where $\alpha $ is the strength of the potential, $u_{j}$ is the position of
the $j$-th plate, and the stable buckled positions are given by $u_{j}=\pm
X_{0}$,\ as illustrated in Fig.\ref{FigBoundary}(b). The local minima of the
potential (\ref{Pote}) are given by 
\begin{equation}
u_{j}=\pm X_{0}.  \label{Mini}
\end{equation}%
We define the normalized plate position by%
\begin{equation}
s_{j}\equiv u_{j}/X_{0}=\pm 1.  \label{sj}
\end{equation}

The potential barrier is given by 
\begin{equation}
U_{0}=\frac{\alpha }{2}X_{0}^{4}.  \label{Pote0}
\end{equation}%
It is necessary to control a buckled plate between two stable positions.

The simplest is to use an electrostatic interaction between the two adjacent
plates as in Fig.\ref{FigBoundary}(a). There is another way to use a
comb-teeth MEMS between the two adjacent plates, about which we discuss in
Appendix \ref{SecCombTeeth}.

In the parallel-plate MEMS, the capacitance between the plate $u_{j}$ and $%
u_{j+1}$ is well described by

\begin{equation}
C_{\text{para}}(u_{j},u_{j+1})=\frac{\varepsilon _{0}S}{X_{\text{cap}%
}+u_{j}-u_{j+1}},  \label{Cu}
\end{equation}%
where $X_{\text{cap}}$ is the distance between the adjacent plates without
buckling, $S$\ is the area of the plate, and $\varepsilon _{0}$ is the
permittivity. The electrostatic potential is given by%
\begin{equation}
U_{\text{cap}}=\sum_{j=1}^{N-1}\frac{C_{\text{para}}(u_{j},u_{j+1})}{2}%
V_{j}^{2},  \label{Ucap}
\end{equation}%
when we control the voltage $V_{j}$\ between the $j$-th and $(j+1)$-th
plates.

\begin{figure}[t]
\centerline{\includegraphics[width=0.42\textwidth]{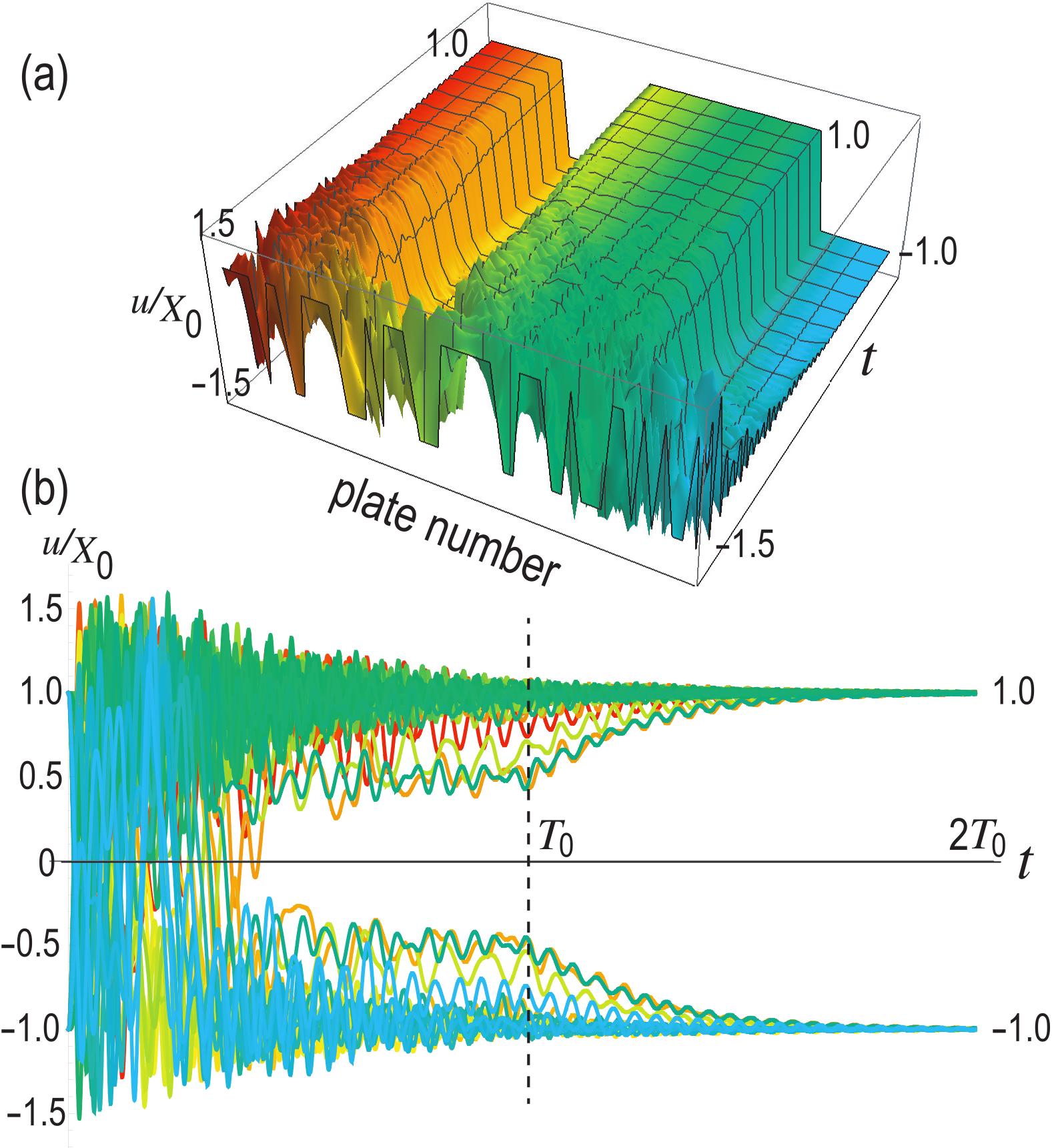}}
\caption{Time evolution of the annealing process under the fixed-boundary
condition, where the exact ground state is achieved as in Fig.\protect\ref%
{FigDynamicsF}(c6). (a) Bird's eye's view and (b) colored plot for each
plate. The vertical axis is the plate position $u/X_{0}$. We have used a
50-plate system, and applied $V_{0}=4V_{\text{cr}}$. The parameters are the
same as in Fig.\protect\ref{FigDynamicsF}. We have used $10^{5}$ time steps
for the calculation.}
\label{FigAnnealF}
\end{figure}

\red{The classical bit information $\pm 1$\ is carried by the variable $u_{j}$%
, from which the standard bit $b_{j}$\ is defined by%
\begin{equation}
u_{j}=\pm X_{0}=\left( -1\right) ^{s_{j}}X_{0},  \label{us}
\end{equation}%
where $b_{j}=0$\ or $1$. The $M$-bit state $\left\vert b_{M}b_{M-1}\cdots
b_{2}b_{1}\right\rangle $\ is uniquely assigned to the digit state $%
\left\vert N\right\rangle $, with $N$\ being defined by%
\begin{equation}
N=\sum_{j=1}^{M}2^{j-1}b_{j}.  \label{BitState}
\end{equation}%
It is sometimes convenient to use $N$\ to represent the $M$-bit state $%
\left\vert b_{M}b_{M-1}\cdots b_{2}b_{1}\right\rangle $.}

\section{Ferromagnetic Ising model}

\subsection{Ising model representation}

We construct an Ising model by considering the buckled-plate MEMS
illustrated in Fig.\ref{FigBoundary}(a). The potential energy of the system
consists of the mechanical part and the electrostatic part, 
\begin{equation}
U_{\text{total}}=U_{\text{mech}}+U_{\text{cap}},  \label{PotenU}
\end{equation}%
where $U_{\text{mech}}$ and $U_{\text{cap}}$\ are given by Eqs.(\ref{Pote})
and (\ref{Ucap}). Provided the applied voltage $V_{j}$ is sufficiently
smaller than a certain critical value $V_{\text{cr}}$, the ground state is
given by Eq.(\ref{Mini}), or $s_{j}=u_{j}/X_{0}=\pm 1$. The critical voltage
is given by%
\begin{equation}
V_{\text{cr}}=\sqrt{\frac{2\alpha X_{0}^{2}\left( X_{\text{cap}%
}^{2}-X_{0}^{2}\right) ^{2}}{3\sqrt{3}\varepsilon _{0}SX_{\text{cap}}}},
\label{CriticalV}
\end{equation}%
as we derive in Appendix \ref{SecCritical}: See Eq.(\ref{MemsCrilV}).

Such a binary system is well described by the Ising model. Indeed, it is
possible to rewrite the total potential energy (\ref{PotenU}) in the form of%
\begin{equation}
U_{\text{Ising}}=\sum_{j=1}^{N-1}J_{j}s_{j}s_{j+1}+E_{0},
\end{equation}%
with the use of the system parameters, 
\begin{equation}
J_{j}=-\frac{2\varepsilon _{0}SV_{j}^{2}X_{0}^{2}}{X_{\text{cap}}\left( X_{%
\text{cap}}^{2}-4X_{0}^{2}\right) },  \label{EqBB}
\end{equation}%
and%
\begin{equation}
E_{0}=\frac{2\varepsilon _{0}SV_{j}^{2}\left( X_{\text{cap}%
}^{2}-X_{0}^{2}\right) }{X_{\text{cap}}\left( X_{\text{cap}%
}^{2}-4X_{0}^{2}\right) },
\end{equation}%
as we derive in Appendix: See Eqs.(\ref{IsingH}), (\ref{EqB}) and (\ref{EqD}%
) in Appendix \ref{Rep}. The system is FM because $J_{j}$\ is negative
definite as in (\ref{EqBB}).

\subsection{Dissipation Mechanism}

We start with an initial state, where $s_{j}=\pm 1$ is randomly assigned to
the site $j$. The configuration is stable without voltage $(V=0)$. We switch
on the voltage beyond the critical value and fix it at $V=V_{0}$. Then,
plates begin to move and the buckling direction may be flipped.

\begin{figure}[t]
\centerline{\includegraphics[width=0.48\textwidth]{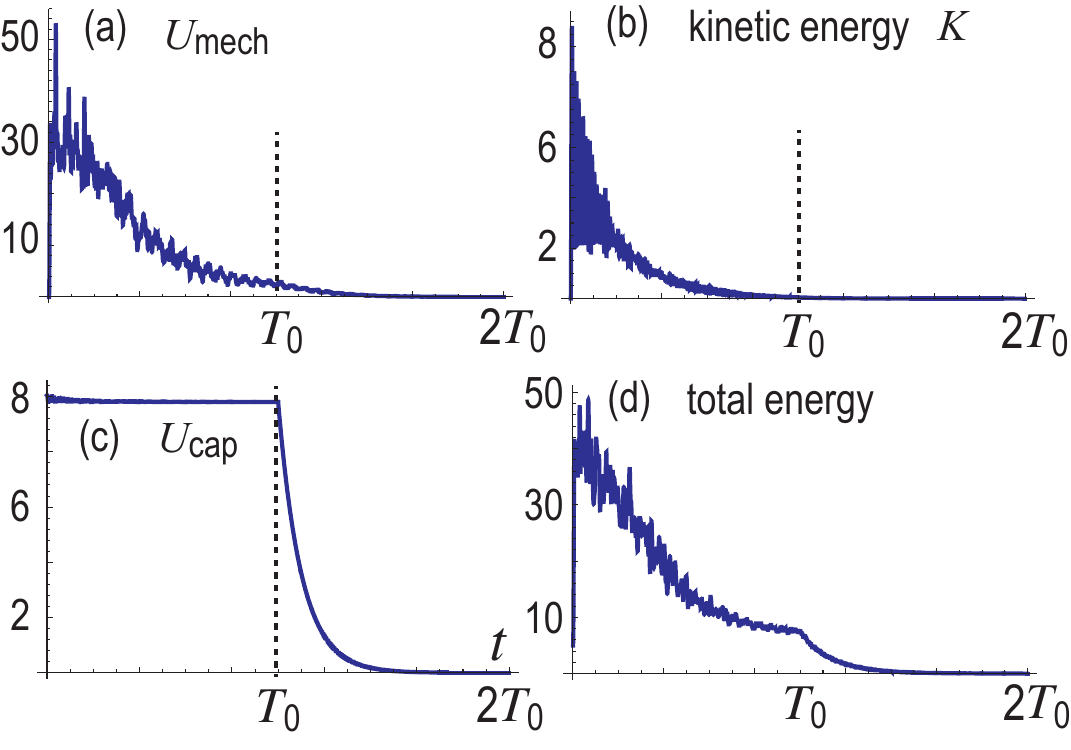}}
\caption{Time evolution of (a) the mechanical potential energy $U_{\text{mech%
}}$, (b) the kinetic energy $K$, (c) the electrostatic potential $U_{\text{%
cap}}$ and (d) total energy under the fixed-boundary condition. We have used
a 16-plate system, and applied $V_0=20V_{\text{cr}}$. The parameters are the
same as in Fig.\protect\ref{FigDynamicsF}. The energy is in the unit of $%
\protect\varepsilon _{0}SV_{0}^{2}/X_{0}$.}
\label{FigEnergyF}
\end{figure}

\begin{figure*}[t]
\centerline{\includegraphics[width=0.98\textwidth]{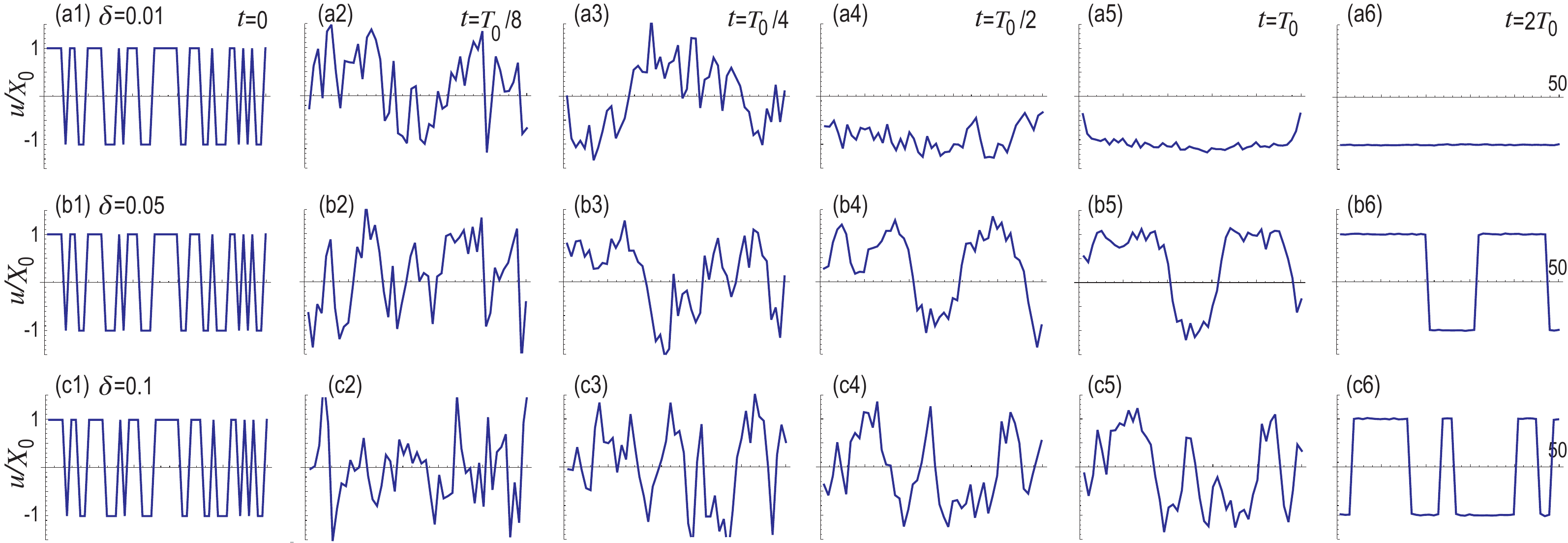}}
\caption{Dynamics of FM interactions under the fixed-boundary condition in
the presence of disorder in $X_{\text{cap}}$. (a1)$\sim $(c1) Configuration $%
u_{j}/X_{0}$ at the initial state $t=0$, which is taken identical to the one
in Fig.\protect\ref{FigDynamicsF}; (a2)$\sim $(c2) that at $t=T_{0}/8$; (a3)$%
\sim $(c3) that at $t=T_{0}/4$; (a4)$\sim $(c4) that at $t=T_{0}/2$; (a5)$%
\sim $(c5) that at $t=T_{0}$; (a6)$\sim $(c6) that at $t=2T_{0}$. (a1)$\sim $%
(a6) The disorder strength is $\protect\delta =0.01$; (b1)$\sim $(b6) it is $%
\protect\delta =0.05$; (c1)$\sim $(c6) it is $\protect\delta =0.1$. (a1)$%
\sim $(a6) The annealing process is perfect when the disorder is weak $%
\protect\delta =0.01$. (b1)$\sim $(b6) There remain three domain walls after
annealing when the disorder is medium $\protect\delta =0.05$. (c1)$\sim $%
(c6) There remain seven domain walls after annealing when the disorder is
large $\protect\delta =0.1$. The vertical axis is the configuration $%
u_{j}/X_{0}$, while the horizontal axis is the position of the plate. We
have used a 50-plate system, and applied $V=20V_{\text{cr}}$. We have set $%
\protect\alpha =0.02$, $\protect\gamma =0.01$, $T_{0}=500$ and $X_{0}=10$. }
\label{FigDisorderF}
\end{figure*}

We study the dynamics of the system. The Lagrangian is given by the kinetic
term $K$ and the potential (\ref{PotenU}),%
\begin{equation}
L=K-U_{\text{total}},
\end{equation}%
with the kinetic energy%
\begin{equation}
K=\sum_{j}\frac{m}{2}\dot{u}_{j}^{2},  \label{Kine}
\end{equation}%
where $\dot{u}_{j}=du_{j}/dt$ is the velocity of the plate. The
Euler-Lagrange-Rayleigh equation is%
\begin{equation}
\frac{d}{dt}\left( \frac{\partial L}{\partial \dot{u}_{j}}\right) -\frac{%
\partial L}{\partial u_{j}}+\frac{\partial R}{\partial \dot{u}_{j}}=0,
\label{EuLaA}
\end{equation}%
where we have introduced the Rayleigh dissipation function in order to
describe the damping of the mechanical motion, 
\begin{equation}
R=\sum_{j}\frac{\gamma }{2}\dot{u}_{j}^{2},  \label{Rayleigh}
\end{equation}%
with $\gamma $ being the damping factor. Note that the external voltage $%
V=V_{0}$\ is just a constant.

The total energy is obtained by using the Lagrangian as%
\begin{equation}
E\equiv \frac{\partial L}{\partial \dot{u}_{j}}\dot{u}_{j}-L.
\end{equation}%
The time dependence of the total energy is given by%
\begin{equation}
\frac{dE}{dt}=-\dot{u}_{j}\frac{\partial R}{\partial \dot{u}_{j}}%
=-\sum_{j}\gamma \dot{u}_{j}^{2},  \label{EneA}
\end{equation}%
where we have used the relation%
\begin{align}
\frac{dL}{dt} =&\dot{q}_{i}\left( \frac{d}{dt}\frac{\partial L}{\partial 
\dot{u}_{j}}+\frac{\partial R}{\partial \dot{u}_{j}}\right) +\frac{\partial L%
}{\partial \dot{u}_{j}}\dot{u}_{ji}  \notag \\
=&\frac{d}{dt}\left( \frac{\partial L}{\partial \dot{u}_{j}}\dot{u}%
_{j}\right) +\dot{u}_{j}\frac{\partial R}{\partial \dot{u}_{j}},
\end{align}%
with the aid of the Euler-Lagrange-Rayleigh equation (\ref{EuLaA}). Eq.(\ref%
{EneA}) means that the total energy monotonically decreases as a function of
time in the presence of the dissipation term. Eventually, the motion of the
plates must stop\ due to dissipation,%
\begin{equation}
\dot{u}_{j}=0.
\end{equation}%
This occurs in the presence of the voltage $V_{0}$.

Finally, we decrease adiabatically the voltage to $V=0$, keeping $\dot{u}%
_{j}=0$. This is a rescaling process, after which a local minimum $u_{j}=\pm
X_{0}$\ or $s_{j}=\pm 1$\ is achieved. This is the basis of the annealing
process in the buckled-plate MEMS system based on a dissipation mechanism.

\subsection{Dynamics of plates}

We shall demonstrate explicitly how the annealing proceeds and how the plate
motion stops to reach a local minimum based on numerical simulations. The
dynamics of plates is determined by Eq.(\ref{EuLaA}), which is explicitly
written down as%
\begin{align}
m\ddot{u}_{j}+\gamma \dot{u}_{j}& =-2\alpha u_{j}\left(
u_{j}^{2}-X_{0}\right) -\frac{\varepsilon _{0}SV_{j}^{2}}{\left( X_{\text{cap%
}}+u_{j}-u_{j+1}\right) ^{2}}  \notag \\
& +\frac{\varepsilon _{0}SV_{j}^{2}}{\left( X_{\text{cap}}+u_{j-1}-u_{j}%
\right) ^{2}}.  \label{EuLaB}
\end{align}%
We solve this equation numerically. We adopt the fixed-boundary condition,
where the outmost plates are fixed without buckling as shown in Fig.\ref%
{FigBoundary}(a). The FM ground state\ is achievable only in this condition.
We apply voltage $0$\ to the ($2n$)-th plate and $V$ to the ($2n+1 $)-th
plate as shown in Fig.\ref{FigBoundary}(a), where the potential difference
between two adjacent plates is $\pm V$. We control $V$\ according to the
formula%
\begin{equation}
V=\left\{ 
\begin{array}{ccc}
V_{0} & \text{for} & t<T_{0} \\ 
V_{0}\exp \left[ -\frac{t-T_{0}}{\tau }\right] & \text{for} & t\geq T_{0}%
\end{array}%
\right. ,
\end{equation}%
where $\tau $ is the decay rate of the voltage.

First, we set an initial state randomly at the two local minima $s_{j}=\pm
1. $ There are a number of domain walls as in Fig.\ref{FigDynamicsF}(a1),
(b1) and (c1). \ When $V_{0}$\ is small enough, the initial state stands as
it is. When it overcomes the potential barrier (\ref{Pote}) of the buckling,
plates begin to move.

Second, we apply a constant voltage $V_{0}$ larger than the critical one $V_{%
\text{cr}}$ given in Eq.(\ref{CriticalV}). We show snapshots of the plate
positions $u_{j}$ for three typical values of $V_{0}$ in Fig.\ref%
{FigDynamicsF}. See also Fig.\ref{FigAnnealF} for a continuous time
evolution in the case of $V_{0}=20V_{\text{cr}}$. As the time evolves, the
number of domain walls decreases to make the electrostatic energy decrease.
We keep the constant voltage $V=V_{0}$ for a while ($0<t<T_{0}$) so that the
flipping of plates is completed, as indicated by a vertical dotted line at $%
t=T_{0}$ in Fig.\ref{FigAnnealF}(b). Here, the position $u_{j}$ deviates
from the local minima $u_{j}=\pm X_{0}$ because of the presence of $V_{0}$.

The final step is the rescaling process, by adiabatically decreasing the
voltage to converge the position $u_{j}$ to the local minimum $u_{j}=\pm
X_{0}$ as in Fig.\ref{FigAnnealF}(b). The FM ground state is achieved when $%
V_{0}$\ is taken large enough as in Fig.\ref{FigDynamicsF}(c6).

It is instructive to see the time evolution of various energies. We show the
time evolution of the mechanical potential energy (\ref{Pote}) in Fig.\ref%
{FigEnergyF}(a). It decreases with oscillations but does not reach the zero
energy for $t<T_{0}$, because the relaxed position deviates from the local
minima (\ref{Mini}) due to the electrostatic force. After the voltage $V$
becomes sufficiently small for $t\gg T_{0}$, the potential energy goes to
zero. On the other hand, the kinetic energy (\ref{Kine}) of the plates
decreases to zero even for $T<T_{0}$ due to the Rayleigh dissipation term (%
\ref{Rayleigh}), as shown in Fig.\ref{FigEnergyF}(b). We also show the time
evolution of the electrostatic potential in Fig.\ref{FigEnergyF}(c). It is
almost constant for $T<T_{0}$ and rapidly decreases for $T>T_{0}$. The time
evolution of the total energy is given in Fig.\ref{FigEnergyF}(d).

This is the annealing process of the present system. The damping of the
mechanical motion plays an essential role in the present annealing process.
It is because the total energy automatically decreases due to the damping
term, and the plate positions are optimized in this process. It is highly
contrasted to the classical and quantum annealing, where it is necessary to
gradually decrease temperature or magnetic field due to the lack of
dissipation mechanism.

\subsection{Effect of randomness}

Next, we study effects of randomness. We introduce randomness into the plate
distance $X_{\text{cap}}$ uniformly distributing from $-\delta $ to $\delta $%
. Then, the distance $X_{\text{cap}}$\ becomes node-dependent, $X_{\text{cap}%
}\rightarrow X_{\text{cap},n}$, where%
\begin{equation}
X_{\text{cap},n}=X_{\text{cap}}\left( 1+\eta _{n}\delta \right) ,
\end{equation}%
with $\eta _{n}$ being a random variable ranging from $-1$ to $1$. We show
the time evolution of the plate positions $u_{j}$ by applying $V_{0}=20V_{%
\text{cr}}$ in Fig.\ref{FigDisorderF}. When the disorder strength is not too
large, the annealing process is perfectly executed as shown in Fig.\ref%
{FigDisorderF}(a6). Hence, it is necessary to make a sample where the
disorder is smaller than $\delta =0.01$ in order to execute annealing
process. We comment that the presence of randomness is not serious since it
is remedied by introducing additional springs to the buckled-MEMS, about
which we discuss in Appendix \ref{Rep}.

\begin{figure}[t]
\centerline{\includegraphics[width=0.48\textwidth]{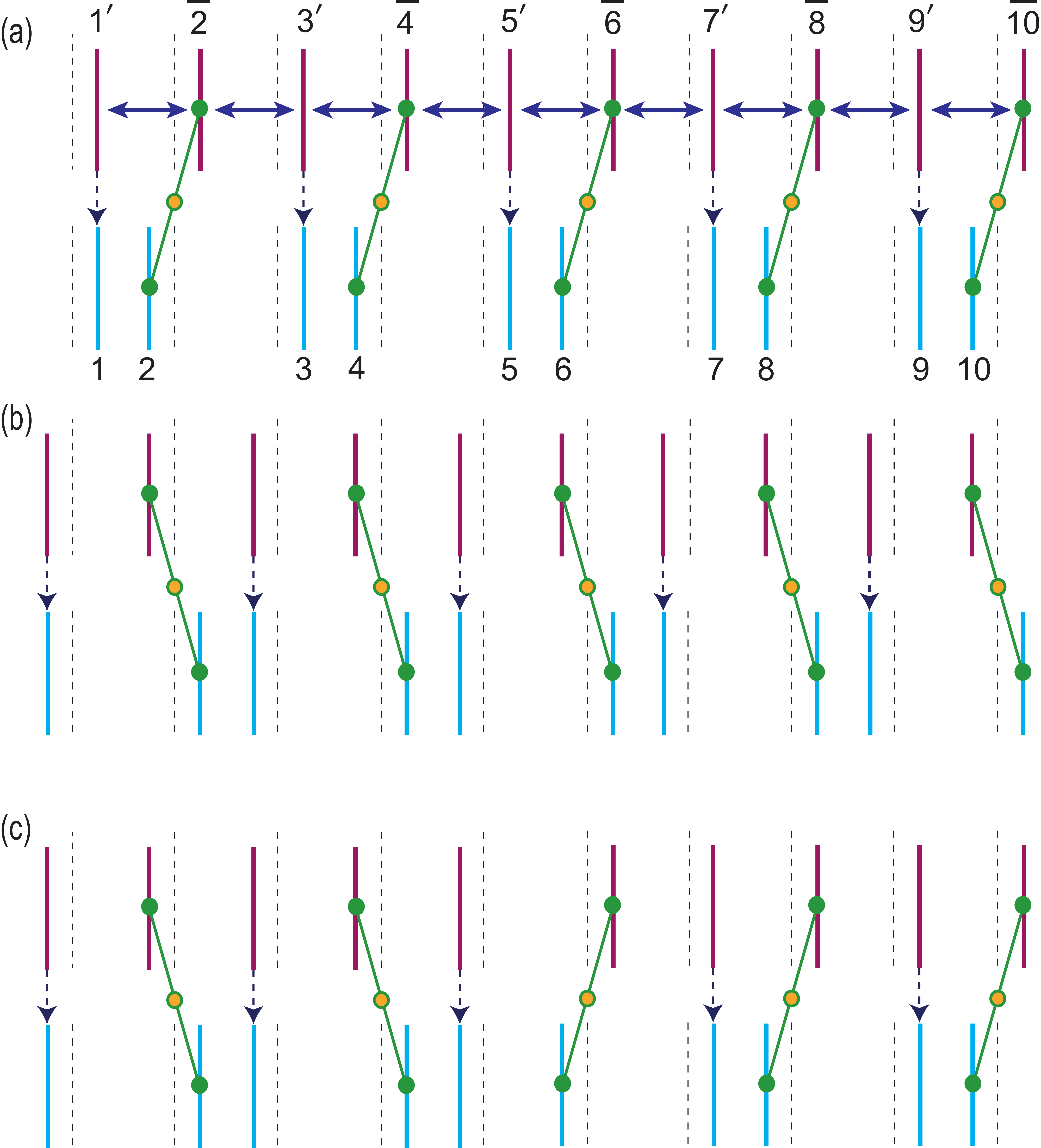}}
\caption{Illustration of a MEMS realizing AF interactions. (a) This is a
ground state representing an AF configuration in the lower chain. \red{%
Double-headed arrows represent electrostatic interactions.} (b) This is the
other AF ground state. The ground states are two-fold degenerate. (c) An
example of excited states, containing one domain wall. \red{Purple plates
are metallic plates and move dynamically by electrostatic interactions,
while cyan plates are plastic plates representing the original bit $b_j$.
There are two types of purple plates: One is connected to a cyan plate by a
seesaw, representing the inverted bit $\bar{b}_j$; the other is connected to
a cyan plate without a seesaw, representing the copied bit $b^{\prime }_j$.
The cyan plate indexed by the original bit $b_j$ 
moves in the forward (reverse) direction of the paired purple plate
indexed by the copied bit $b^{\prime }_j$ (inverted bit $\bar{b}_j$). 
The AF order is realized for the original
bits in the lower chain in this way. }}
\label{FigAF}
\end{figure}

\begin{figure*}[t]
\centerline{\includegraphics[width=0.98\textwidth]{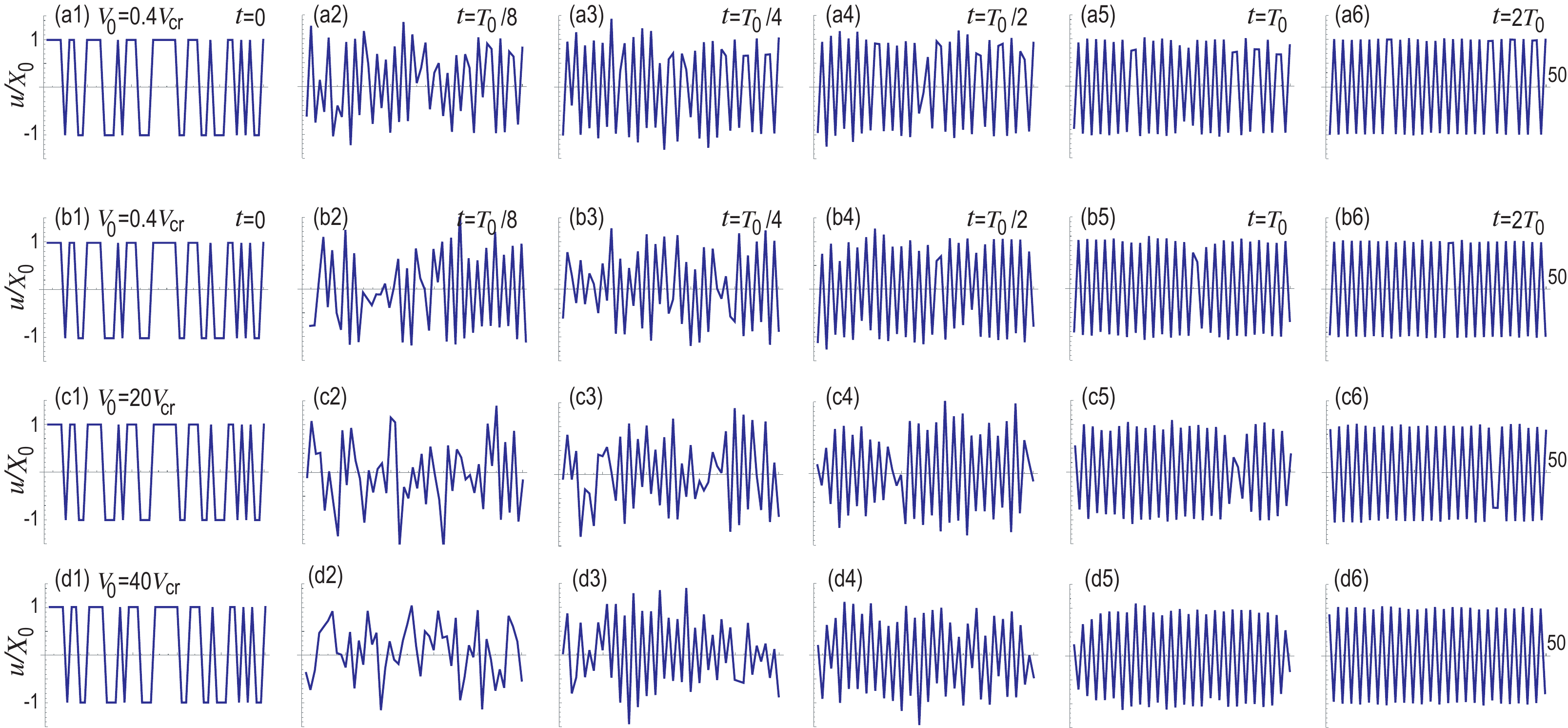}}
\caption{ Dynamics of AF interactions with the fixed-boundary condition. (a1)%
$\sim $(c1) Configuration $u_{j}/X_{0}$ at the initial state $t=0$; (a2)$%
\sim $(c2) that at $t=T_{0}/8$; (a3)$\sim $(c3) that at $t=T_{0}/4$; (a4)$%
\sim $(c4) that at $t=T_{0}/2$; (a5)$\sim $(c5) that at $t=T_{0}$; (a6)$\sim 
$(c6) that at $t=2T_{0}$. We have used a 50-plate system. (a1)$\sim $(a6)
The applied voltage is $V=0.4V_{\text{cr}}$; (b1)$\sim $(b6) it is $V=4V_{%
\text{cr}}$; (c1)$\sim $(c6) it is $V=20V_{\text{cr}}$ and (d1)$\sim $(d6)
it is $V=40V_{\text{cr}}$. (a1)$\sim $(a6) There is no annealing when the
voltage is below the critical value ($V<V_{\text{cr}}$). (b1)$\sim $(b6) and
(c1)$\sim $(c6) There remain some domain walls after annealing when the
voltage exceeds the critical value but not sufficiently. (d1)$\sim $(d6) The
annealing process is perfect when the voltage exceeds the critical value
sufficiently. The vertical axis is the configuration $u_{j}/X_{0}$. The
horizontal axis is the position $j$. We have set $\protect\alpha =0.02$, $%
\protect\gamma =0.01$, $T_{0}=500$ and $X_{0}=10$. }
\label{FigAFDynamics}
\end{figure*}

\section{Antiferromagnetic Ising model}

\label{SecAnti}

\red{In the above construction, only the FM interactions are realized.
However, AF interactions are also necessary in order to solve nontrivial
Ising problems. We propose a seesaw mechanism shown in Fig.\ref{FigAF}(a),
where an even-numbered plate (in purple) in the upper chain is connected to
a plate (in cyan) in the lower chain by a spring (in green) rotating around
an fixed point (in orange). These two plates move simultaneously preserving
the condition%
\begin{equation}
u_{j}\equiv \left( -1\right) ^{j}u_{j}^{\text{lower}}.  \label{EqU}
\end{equation}%
On the other hand, an odd-numbered plate (in purple) in the upper chain
moves precisely in the same way as a plate (in cyan) in the lower chain. The
electrostatic interaction operates between the adjacent plates only in the
upper chain. Plates in the lower chain are made of nonmetallic material such
as plastic and gum, and are not components of MEMS.}

\red{When the bit $b_{j}$\ assigned in the lower chain and the bit $\overline{b}%
_{j}$\ in the upper chain are connected by a seesaw, the condition (\ref{EqU}%
) introduces the relation between them,%
\begin{equation}
\overline{0}_{j}=1_{j},\qquad \overline{1}_{j}=0_{j}.  \label{InvBit}
\end{equation}%
The barred bit is called the inverted bit. On the other hand, when they are
not connected by a seesaw, the bits are identical between the two chains,%
\begin{equation}
b_{j}^{\prime }=b_{j}.  \label{CopyBit}
\end{equation}%
The dashed bit is called the copied bit.} 

We now argue that the AF Ising model is realized in the lower chain in Fig.%
\ref{FigAF}. By inserting Eq.(\ref{EqU}) to the Hamiltonian (\ref{Hs}), we
obtain\ the Hamiltonian 
\begin{align}
H_{j}\left( s_{j},s_{j+1}\right) =& -J_{j}s_{j}s_{j+1}+\frac{B_{j}}{2}\left(
-1\right) ^{j}s_{j}  \notag \\
& +\frac{B_{j+1}}{2}\left( -1\right) ^{j+1}s_{j+1}+\frac{E_{0}}{N-1},
\end{align}%
which represents the AF Ising model. The electrostatic potential is
represented as%
\begin{equation}
U_{\text{cap}}=\sum_{j=1}^{N-1}\frac{C_{\text{para}}(\left( -1\right)
^{j}u_{j},\left( -1\right) ^{j+1}u_{j+1})}{2}V_{j}^{2},
\end{equation}%
in terms of the position $u_{j}$ of the plate in the lower chain, where the
use was made of Eq.(\ref{EqU}). By inserting it to the
Euler-Lagrange-Rayleigh equation (\ref{EuLaA}), we obtain the equations of
motion for the lower chain%
\begin{align}
m\left( -1\right) ^{j}\ddot{u}_{j}& +\gamma \left( -1\right) ^{j}\dot{u}_{j}
\notag \\
=& -2\beta \left( -1\right) ^{j}u_{j}\left( u_{j}^{2}-X_{0}\right)  \notag \\
& -\frac{\varepsilon _{0}SV_{j}^{2}}{\left( X_{\text{cap}}+\left( -1\right)
^{j}u_{j}-\left( -1\right) ^{j+1}u_{j+1}\right) ^{2}}  \notag \\
& +\frac{\varepsilon _{0}SV_{j}^{2}}{\left( X_{\text{cap}}+\left( -1\right)
^{j-1}u_{j-1}-\left( -1\right) ^{j}u_{j}\right) ^{2}}.
\end{align}

We show numerical results of the dynamics in the configuration in Fig.\ref%
{FigAFDynamics} for various applied constant voltage $V_{0}$. When we apply
a weak voltage, there is almost no annealing and there are many domain walls
as in Fig.\ref{FigAFDynamics}(a6). When we apply a medium voltage, the
annealing process is not sufficient and the final state have AF domain walls
in an instance of Fig.\ref{FigAFDynamics}(b6) and (c6). When we apply a
strong voltage, the final state is perfectly antiferromagnet as an instance
of Fig.\ref{FigAFDynamics}(d6). These results show that AF interactions are
well simulated by using the seesaw configuration shown in Fig.\ref{FigAF}.

\begin{figure*}[t]
\centerline{\includegraphics[width=0.8\textwidth]{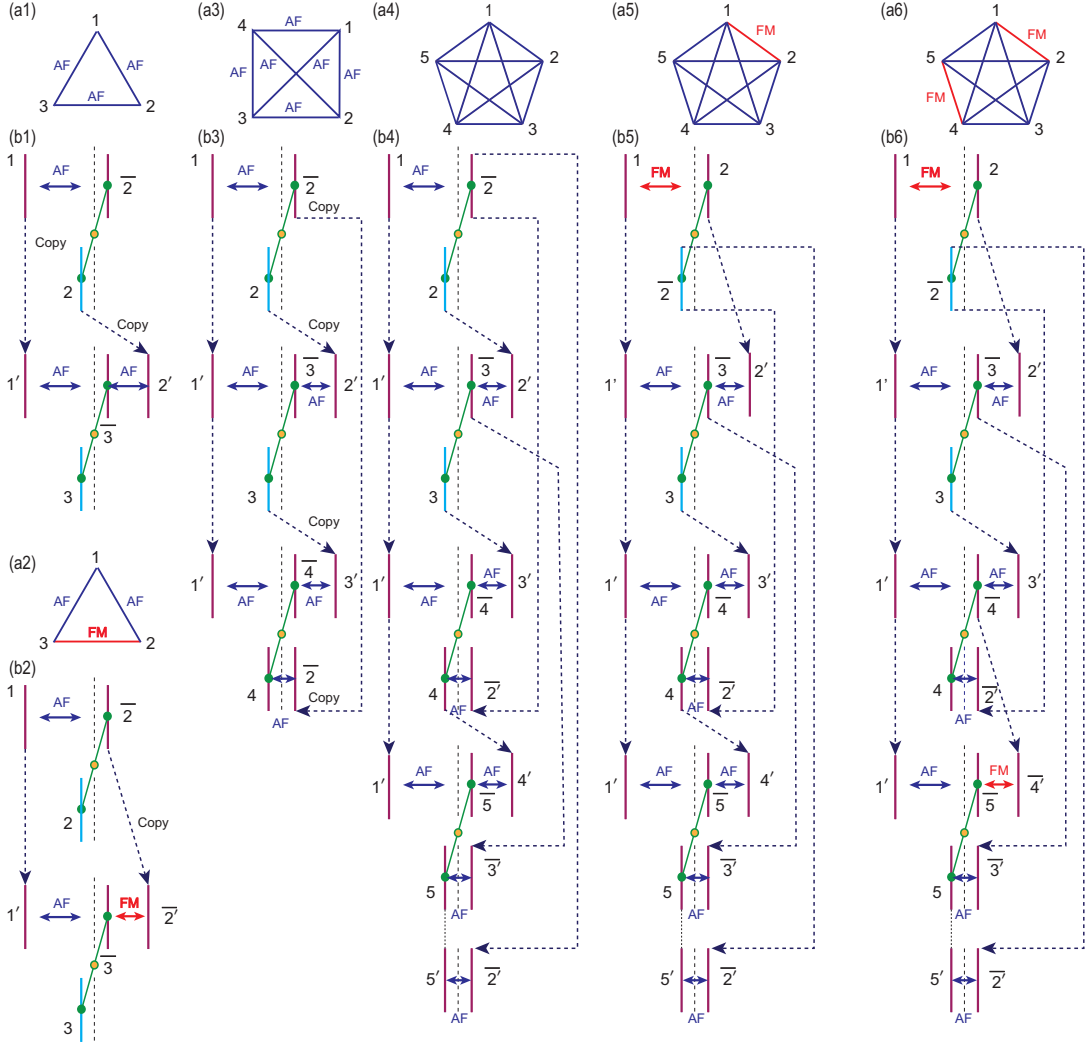}}
\caption{ \red{(a) Illustration of fully-connected bits. Ferromagnetic
interactions are marked in red lines and AF interactions are marked in blue
lines. (a1) Fully-connected three bits with AF interactions. (a2)
Fully-connected three bits with two AF and one FM interactions. (a3)
Fully-connected four bits with AF interactions. (a4) Fully-connected five
bits with AF interactions. (a5) Fully-connected five bits with AF
interactions except for one FM interaction. (a6) Fully-connected five bits
with AF interactions except for two FM interactions. (b1)$\sim$(b6)
Illustration of a fully-connected buckled-plate MEMS. Effective
electrostatic interactions are indicated by double-headed arrows between two
purple plates. The number in the vicinity of a plate indicates the number of
a bit. A pair of plates connected by a dotted arrow are indexed by the
original bit $b_j$ and the copied bit $b_{j}^{\prime }$, and they move
simultaneously. A pair of plates connected by a seesaw are indexed by the
original bit $b_{j}$ and the inverted bit $\overline{b_{j}}$, and they move
inversely. }}
\label{FigFull}
\end{figure*}

\begin{figure*}[t]
\centerline{\includegraphics[width=0.88\textwidth]{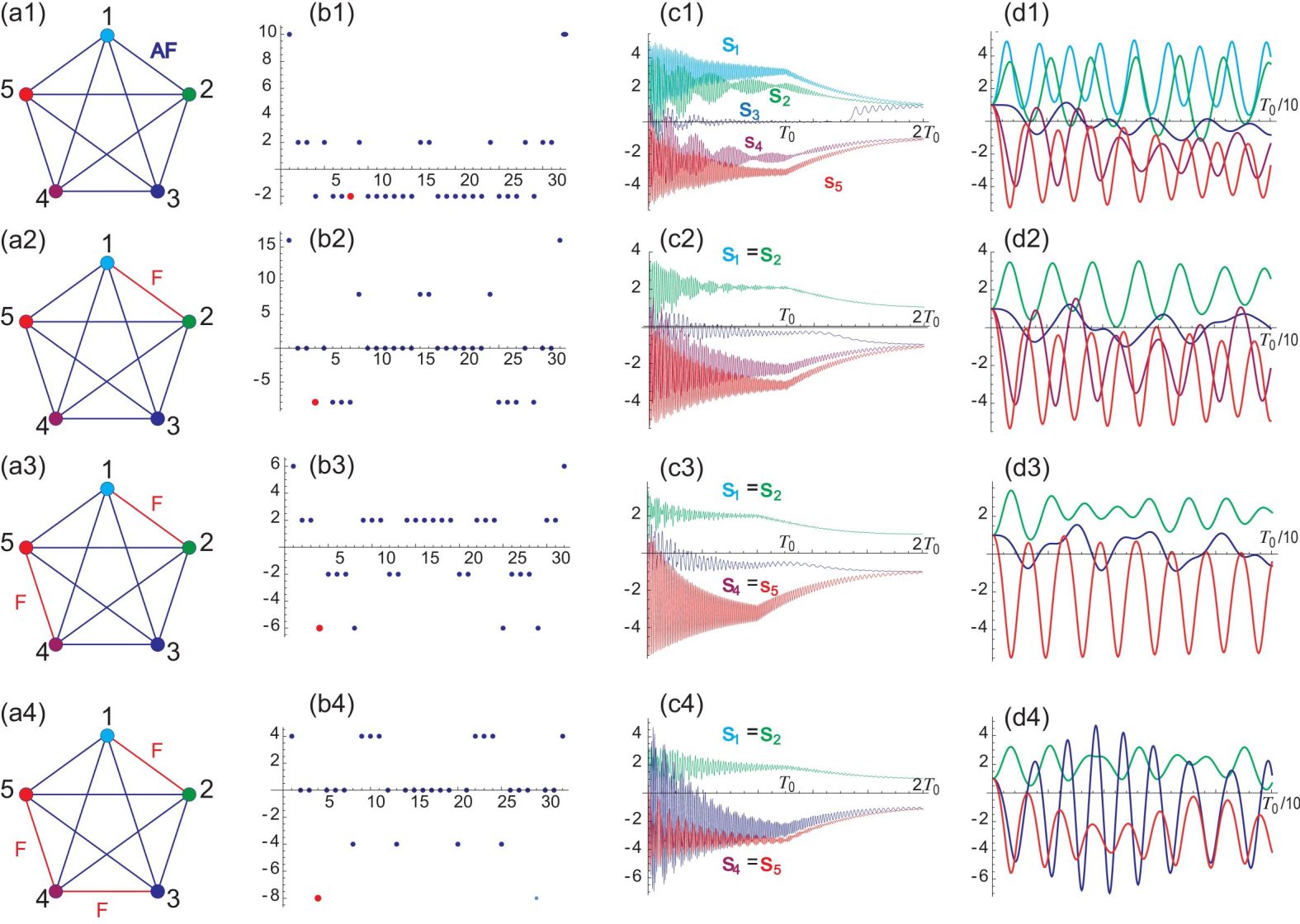}}
\caption{ \red{(a) Illustration of fully-connected five bits, which are the same
as in Fig.\protect\ref{FigFull}(a). (b) The energy spectrum as a function of
the digit state $\left\vert N\right\rangle $ defined by Eq.(\protect\ref%
{BitState}). The numerical results are marked in red, which is one of the degenerated
ground states. When we start from the initial state $\left\vert
00000\right\rangle =\left\vert 0\right\rangle $, one of the ground state $%
\left\vert 00111\right\rangle =\left\vert 7\right\rangle $ is chosen in
(b1), while $\left\vert 00011\right\rangle =\left\vert 3\right\rangle $ is
chosen in (b2), (b3) and (b4) by the annealing process. The vertical axis is
the energy in units of $J$. (c) Time evolution of the annealing process for $%
0\leq t\leq 2T_{0}$, where the vertical axis stands for the normalized plate
position $s_{j}$. The first normalized plate position $s_{1}$ is indicated
by cyan curve, the second one $s_{2}$ by green curve, the third
one $s_{3}$ by blue curve, the fourth one $s_{4}$ by purple curve and the fifth one $s_{5}$ by red curve. We have
set $\protect\alpha =0.02$, $\protect\gamma =0.01$, $T_{0}=500 $ and $%
X_{0}=10$. (d) Initial time evolution of the annealing process for $0\leq
t\leq 0.1T_{0}=50$, where the vertical axis is $u_{j}$. We have applied $%
V=40V_{\text{cr}}$.}}
\label{FigFullDynamics}
\end{figure*}

\section{Fully-connected annealing machine}

\red{In order to solve a nontrivial problem, it is necessary to construct a
fully-connected system, where both FM and AF interactions coexist. It looks
that it is almost impossible to construct a fully-connected system because
we need $\left( M-1\right) $ plates interacting electrostatically with one
plate in the present buckled-plate MEMS configuration. However, it has been
proposed in the context of quantum annealing machines that the
fully-connected system can be constructed by introducing copied bits\cite%
{Choi,Oku}.}

\red{We use this idea in the MEMS annealing machine as illustrated in Fig.\ref%
{FigFull}. We prepare a copy of a plate in such a way that the motion is
identical to the original plate. We denote the copied bit of the original
bit $b_{j}$\ by $b_{j}^{\prime }$. It is materialized by bridging two
plates, where we adjust that the total mass of the original and copied bits
is identical. It is always possible because the original and copied bit form
a rigid body.\ Then, we introduce an electrostatic interaction between an
original plate and a copied plate. By adding an appropriate number of copied
plates, it is possible to materialize a fully-connected system. Furthermore,
it is possible to materialize a fully-connected system containing both FM
and AF interactions in a single network by appropriately choosing the
position of interactions.}

\red{We note that the interaction is FM between $b_{j}$\ and $b_{k}$, between $%
b_{j}^{\prime }$\ and $b_{k}$, and between $b_{j}^{\prime }$\ and $%
b_{k}^{\prime }$.\ It is also FM between $\overline{b}_{j}$\ and $\overline{b%
}_{k}$, and so on. On the other hand, the interaction is AF between $b_{j}$\
and $\overline{b}_{k}$, and so on.}

\red{We present how to construct the MEMS annealing machine by taking an instance
of the three-bit fully-connected AF system in Fig.\ref{FigFull}(a1). (i) We
introduce an AF interaction between the bit $1$\ and the inverted bit $%
\overline{2}$ as in the case of the left-most two bits in Fig.\ref{FigAF}.\
(ii) In the same way, we introduce an AF interaction between the bit $1$\
and the inverted bit $\overline{3}$. (iii) As the final step, we introduce
an AF\ interaction between the copied bit $2^{\prime }$ and the inverted bit 
$\bar{3}$\ as in Fig.\ref{FigFull}(b1).}

\red{The next instance is the three-bit fully-connected AF and FM system shown in
Fig.\ref{FigFull}(a2). The first two steps (i) and (ii) are identical. Only
the final step (iii) is different, in which we introduce an FM interaction
between the inverted copied bit $\overline{2}^{\prime }$ and the inverted
bit $\bar{3}$\ as in Fig.\ref{FigFull}(b2).}

\red{This method is applicable to any fully-connected network. In the $M$-bit
system, we introduce $M-1$\ AF links starting from the bit $1$\ by
introducing $M-2$\ copied bits $1^{\prime }$. Then, we introduce an AF or FM
link between any pair of bits except the bit $1$\ by introducing necessary
copied bits. Explicit examples are given in Figs.\ref{FigFull}(b3)\symbol{126%
}(b6) for $M=4$ and $5$.}

\red{For the FM interaction between $j$ and $k$, the equations of motion are
given by%
\begin{align}
m\ddot{u}_{j}+\gamma \dot{u}_{j}+2\alpha u_{j}\left( u_{j}^{2}-X_{0}\right)
& =-\frac{\varepsilon _{0}SV_{jk}^{2}}{\left( X_{\text{cap}%
}+u_{j}-u_{k}\right) ^{2}},  \notag \\
m\ddot{u}_{k}+\gamma \dot{u}_{k}+2\alpha u_{k}\left( u_{k}^{2}-X_{0}\right)
& =-\frac{\varepsilon _{0}SV_{jk}^{2}}{\left( X_{\text{cap}%
}+u_{k}-u_{j}\right) ^{2}}.
\end{align}%
On the other hand, the AF interaction between $j$ and $k$, the equations of
motion are given by%
\begin{align}
m\ddot{u}_{j}+\gamma \dot{u}_{j}+2\alpha u_{j}\left( u_{j}^{2}-X_{0}\right)
& =\frac{\varepsilon _{0}SV_{jk}^{2}}{\left( X_{\text{cap}%
}+u_{j}-u_{k}\right) ^{2}},  \notag \\
m\ddot{u}_{k}+\gamma \dot{u}_{k}+2\alpha u_{k}\left( u_{k}^{2}-X_{0}\right)
& =-\frac{\varepsilon _{0}SV_{jk}^{2}}{\left( X_{\text{cap}%
}+u_{j}-u_{k}\right) ^{2}}.
\end{align}%
More detailed equations are given in Appendix G.}

\red{We present numerical results for five-bit fully-connected networks in Figs.%
\ref{FigFullDynamics}(a1)\symbol{126}(a4). The energy spectrum is shown in
Figs.\ref{FigFull}(b1)\symbol{126}(b4) as a function of the digit state $%
\left\vert N\right\rangle $\ representing the $5$\ bits $\left\vert
b_{5}b_{4}b_{3}b_{2}b_{1}\right\rangle $: See Eq.(\ref{BitState}). For
example, the energy spectrum for this system is shown in Fig.\ref%
{FigFullDynamics}(b1), where more than half of states form the degenerated
ground states. In Fig.\ref{FigFull}(a5), one AF interaction is replaced by a
FM interaction. The corresponding energy spectrum is shown in Fig.\ref%
{FigFullDynamics}(b2), where the degeneracy becomes smaller than the totally
AF system. The dynamics are identical between the bit 1 and the bit 2, which
would be due to the introduction of the FM\ interactions between the bit 1
and the bit 2 as shown in Fig.\ref{FigFullDynamics}(c2). A further
introduction of the FM interactions as shown in Fig.\ref{FigFull}(a6) leads
to a further decrease of the degeneracy of the ground states as shown in Fig.%
\ref{FigFullDynamics}(b4). The dynamics are identical between the bit 4 and
bit 5 in addition to the bit 1 and the bit 2, which would be due to the
introduction of the FM\ interactions between the bit 4 and the bit 5 as
shown in Fig.\ref{FigFullDynamics}(c3). When there are three FM interactions
as in Fig.\ref{FigFullDynamics}(a4), there are only two-fold degenerate
ground states as shown in Fig.\ref{FigFullDynamics}(b4). Nevertheless, one
of the ground states is correctly obtained as in Fig.\ref{FigFullDynamics}%
(c4).}

\red{In Fig.\ref{FigFullDynamics}(c), the numerically solved annealing process
for the 5 qubit system is presented, where we start with the initial
condition $\left\vert 00000\right\rangle =\left\vert 0\right\rangle $. We
find that the exact ground state is reached by the annealing process for
various systems. In all cases, the obtained ground state is $\left\vert
00111\right\rangle =\left\vert 7\right\rangle $ for Fig.\ref{FigFullDynamics}%
(b1) and $\left\vert 00011\right\rangle =\left\vert 3\right\rangle $ for Fig.%
\ref{FigFullDynamics}(b2)\symbol{126}(b4). It indicates that the present
MEMS annealing machine can solve arbitrary nontrivial problems.}

\section{\textbf{Discussions}}

We have proposed an application of MEMS to an Ising machine. \red{In order to
solve nontrivial Ising problems, it is necessary to construct the
Edwards-Anderson model\cite{Edwards}, where the FM and AF interactions
coexist arbitrarily. We have proposed a method to construct a mechanism
solving such a problem with the aid of a fully-connected MEMS\ network.} 

We have used the buckled-plate MEMS to explain an Ising machine for its
simplicity. Actually, to avoid the pull-in instability, an Ising machine
based on the comb-teeth MEMS would be better, about with we discuss in
Appendix \ref{SecDuality}. The comb-teeth MEMS looks a bit complicated but
its fabrication is possible in the present technology. The translation of
mathematical formulas is simple between these two types of MEMS because
there holds a duality relation between them, as we explain in Appendix \ref%
{SecDuality}.

We comment that there is another proposal on the Ising machine based on MEMS%
\cite{Mahb}, where bistable vibrations are used based on the parametric
resonances of multiple modes. It presents a different realization of the
MEMS Ising machine from the present proposal.

In general, temperature is controlled in classical annealing, while external
magnetic field is controlled in quantum annealing. They are set above the
critical values in the initial stage. Then, they are gradually decreased,
during which the annealing proceeds. On the other hand, we control external
voltage in the annealing method based on the buckled-plate MEMS. In the
first step of the annealing process, we keep a constant voltage for a
certain period. The essential part of the annealing is completed due to the
Rayleigh dissipation term. Although we gradually decrease the voltage, this
is just to rescale the position to $u_{j}=\pm X_{0}$. Thus, the annealing
process is quite different from the standard ones. We would like to point
out that the control of the voltage is easier than the control of the
temperature or the magnetic field, which will be of benefit to future
applications. In addition, our mechanism works at the room temperature
without applying external magnetic field.

The speed and energy consumption depend on the size of MEMS. In general, the
speed is higher and the energy consumption is lower if the size of MEMS is
smaller. The flipping speed is around 1kHz in standard MEMS. In the
annealing process shown in Fig.\ref{FigDynamicsF}, the number of flips are
of the order of one thousand times, which requires one second. In this
context, it is better to construct an annealing machine in
Nano-Electro-Mechanical System (NEMS). See Appendix \ref{SecFreq} for the
characteristic frequency.

The speed of our annealing machine based on MEMS may not depend on the
number of the plates because the plates move simultaneously to minimize the
energy. It is contrasted to the digital simulations, where the calculation
is done sequentially in a software model. The speed may be much faster in
the MEMS machine when the number of plates is large. It is a merit of a
machine based on the rules of nature in the spirit of Feynman's idea\cite%
{Feynman} as pointed out in Introduction.

In passing, we would like to report that we have already fabricated some
basic elements of the present MEMS Ising machine\cite{MitaA,MitaB} after
submission of this paper.

M. E. is very much grateful to E. Saito and N. Nagaosa for helpful
discussions on the subject. Y. M. is supported by CREST, JST (JPMJCR20T2).
M. E. is supported by the Grants-in-Aid for Scientific Research from MEXT
KAKENHI (Grants No. JP17K05490 and No. JP18H03676) and CREST, JST
(JPMJCR16F1 and JPMJCR20T2).

\appendix

\section{MEMS memory}

\label{SecMemory}

\begin{figure}[t]
\centerline{\includegraphics[width=0.48\textwidth]{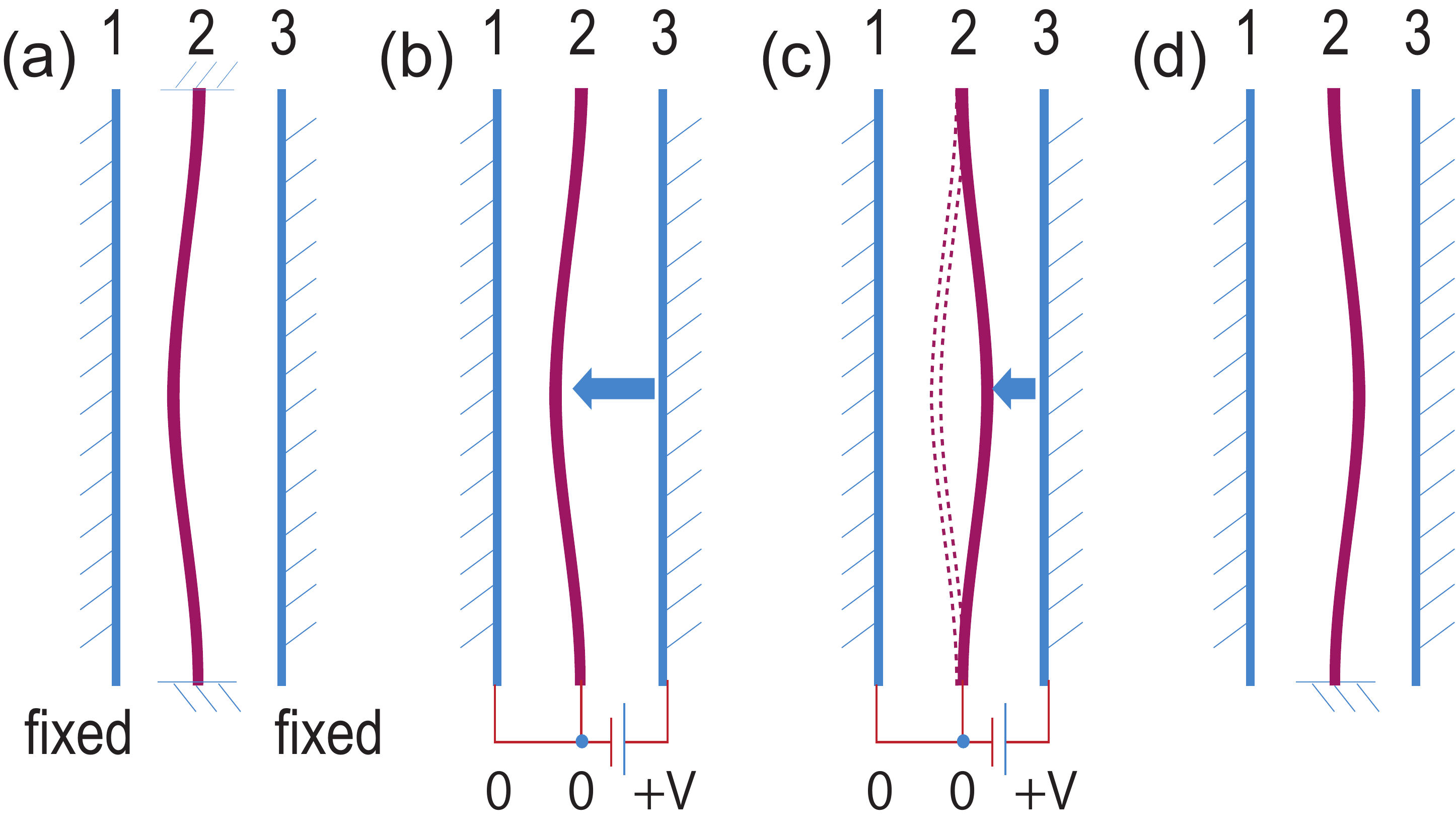}}
\caption{Flip process of the MEMS memory. (a) The initial position of the
plate is $u=-X_{0}$. (b) We apply voltage between the central and the right
plates. (c) The position is flipped from $u=-X_{0}$ to $u=X_{0}$. (d) We
remove the voltage, where the plate is fixed at $u=X_{0}$.}
\label{FigFlip}
\end{figure}

Before we derive the critical voltage of the MEMS Ising machine, i.e., Eq.(%
\ref{CriticalV}) in the main text, we analyze a similar but simpler system,
that is a MEMS memory\cite{Rincon} consisting of three plates as shown in
Fig.\ref{FigFlip}. The outermost two plates are fixed at the position $%
u_{1}=u_{3}=0$, while the inner plate $u\equiv u_{2}$ can move freely. There
are two stable positions $u=\pm X_{0}$, which can store a binary
information. The position of the plate can be read out by measuring the
capacitance. This can be used as a binary memory.

We discuss how to flip the stored information. We choose the initial state $%
u=-X_{0}$, as shown in Fig.\ref{FigFlip}(a). We apply a voltage between the
right and middle plates as shown in Fig.\ref{FigFlip}(b). The electrostatic
potential is given by%
\begin{equation}
C_{\text{para}}(u)=\frac{\varepsilon _{0}S}{X_{\text{cap}}+u}.
\end{equation}%
The static energy is given by%
\begin{equation}
U_{\text{cap}}=\frac{C_{\text{para}}(u)}{2}V^{2},  \label{Ene3}
\end{equation}%
when we apply a constant voltage $V$ between the right and middle plates.

\begin{figure}[t]
\centerline{\includegraphics[width=0.48\textwidth]{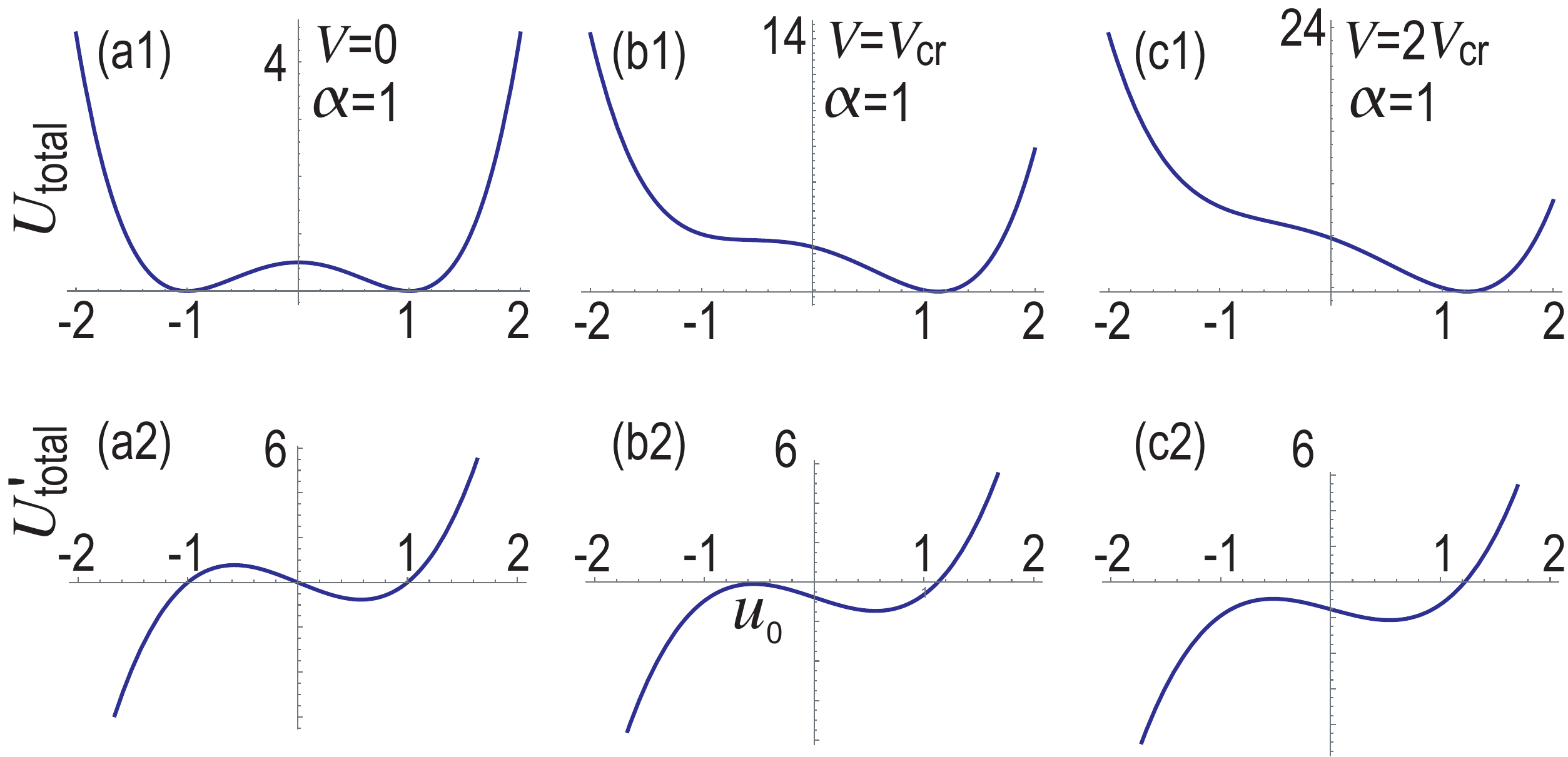}}
\caption{(a1) Total potential $U_{\text{total}}$ with $V=0$ (i.e., the
mechanical potential $U_{\text{mech}}$)$\ $for three plates given by Fig.%
\protect\ref{FigFlip}, where $u_{1}=u_{3}=0$. (b1) $U_{\text{total}}$ with $%
V=V_{\text{cr}}$. (c1) $U_{\text{total}}$ with $V>V_{\text{cr}}$. (a2), (b2)
and (c2) corresponding differential of the potential $U_{\text{total}%
}^{^{\prime }}=\partial U_{\text{total}}/\partial u$. We have set $\protect%
\alpha =1$. The horizontal axis is the position in units of $X_{0}$.}
\label{FigTriPote}
\end{figure}

The position $u$ is determined by the total potential,%
\begin{equation}
U_{\text{total}}(u)=\frac{\alpha }{2}\left( u^{2}-X_{0}^{2}\right) ^{2}+%
\frac{C_{\text{para}}(u)}{2}V^{2}.  \label{UTot}
\end{equation}%
We analyze the condition for the central plate to move from the left
position to the right position without a barrier: See Fig.\ref{FigFlip}(c)
and Fig.\ref{FigTriPote}. The condition is that the slope of the curve $U_{%
\text{total}}(u)$ is negative, $\partial U_{\text{total}}/\partial u<0$, for
all $\left\vert u\right\vert <X_{0}$. In order to solve this condition, we
determine a position $u_{0}$ where the slope of $\partial U_{\text{total}%
}/\partial u$ is zero, or%
\begin{equation}
\left. \frac{\partial ^{2}U_{\text{total}}}{\partial u^{2}}\right\vert
_{u_{0}}=0.  \label{Uuu}
\end{equation}%
If the condition 
\begin{equation}
\left. \frac{\partial U_{\text{total}}}{\partial u}\right\vert _{u_{0}}<0
\label{EqE}
\end{equation}%
is satisfied at this position $u=u_{0}$, it follows that $\partial U_{\text{%
total}}/\partial u<0$ for $u<X_{0}$. See Fig.\ref{FigTriPote}(c2).

By using an approximation $X_{\text{cap}}\gg u$, the static energy is
approximated as%
\begin{equation}
U_{\text{cap}}\simeq \frac{\varepsilon _{0}SV^{2}}{2X_{\text{cap}}}\left( 1-%
\frac{u}{X_{\text{cap}}}\right) .
\end{equation}%
By inserting it to Eq.(\ref{UTot}) and Eq.(\ref{Uuu}), we find%
\begin{equation}
\frac{\partial ^{2}U_{\text{total}}}{\partial u^{2}}=4\alpha u^{2}+2\alpha
\left( u^{2}-X_{0}^{2}\right) =0,
\end{equation}%
whose solution is $u_{0}=-X_{0}/\sqrt{3}$. Now, the condition (\ref{EqE})
yields 
\begin{eqnarray}
\left. \frac{\partial U_{\text{total}}}{\partial u}\right\vert _{u_{0}}
&=&\left. 2\alpha u\left( u^{2}-X_{0}^{2}\right) -\frac{\varepsilon _{0}S}{%
2X_{\text{cap}}^{2}}V^{2}\right\vert _{u_{0}}  \notag \\
&=&\frac{4\alpha X_{0}^{3}}{3\sqrt{3}}-\frac{\varepsilon _{0}S}{2X_{\text{cap%
}}^{2}}V^{2}<0,
\end{eqnarray}%
from which we obtain $V>V_{\text{cr}}$ with%
\begin{equation}
V_{\text{cr}}=\sqrt{\frac{8\alpha X_{0}^{3}X_{\text{cap}}^{2}}{3\sqrt{3}%
\varepsilon _{0}S}}.  \label{CriV}
\end{equation}%
The plate position flips from $u=-X_{0}$\ to $u=X_{0}$\ for $V>V_{\text{cr}}$%
. Finally, we remove the voltage, where the plate is fixed at $u=X_{0}$,\ as
shown in Fig.\ref{FigFlip}(d).

\section{Flipping dynamics}

\label{SecCritical}

\begin{figure}[t]
\centerline{\includegraphics[width=0.48\textwidth]{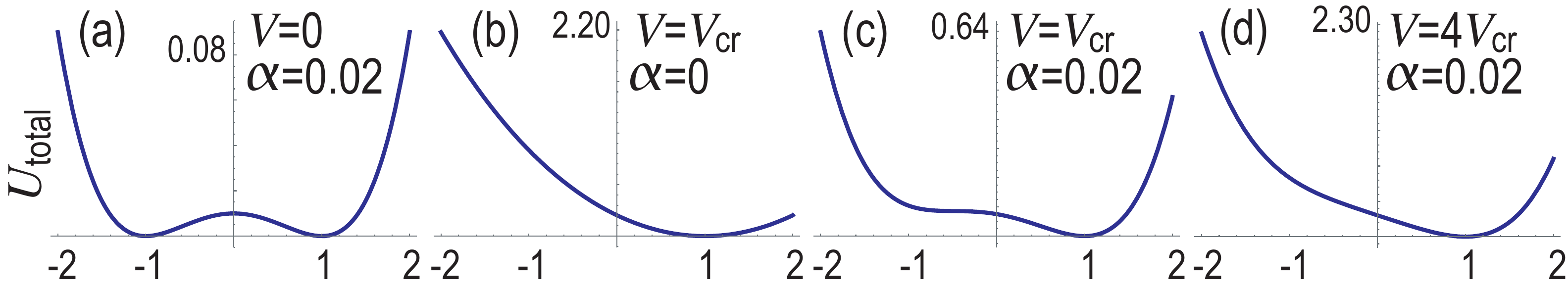}}
\caption{Total potential energy $U_{\text{total}}$. (a) Spring
potential (\protect\ref{Pote}) with $\protect\alpha =0.02$. (b)
Electrostatic potential (\protect\ref{Ucap}) with $\protect\alpha =0$ and $%
V=V_{\text{cr}}$. (c) $\protect\alpha =0.02$ and $V=V_{\text{cr}}$. (d) $%
\protect\alpha =0.02$ and $V=4V_{\text{cr}}.$}
\label{FigPotential}
\end{figure}

The flipping dynamics of the MEMS Ising machine can be understood
intuitively by plotting the potential $U_{\text{total}}=U_{\text{mech}}+U_{%
\text{cap}}$. For simplicity, we first consider three plates, where two
outermost plates are fixed at the position $u_{1}=u_{3}=X_{0}$\ while the
inner plate moves freely. The electrostatic potential is given by%
\begin{equation}
U_{\text{cap}}=\frac{\varepsilon _{0}S}{X_{\text{cap}}-u+X_{0}}+\frac{%
\varepsilon _{0}S}{X_{\text{cap}}-X_{0}+u}.
\end{equation}%
We show the electrostatic potential as a function of the position of the
inner plate $u$\ in Fig.\ref{FigPotential}(b). It takes a minimum at $%
u=X_{0} $, which indicates that the inner plate tends to take the same
position as the outermost plates. There are two minima in the spring
potential energy as shown in Fig.\ref{FigPotential}(a). By increasing the
voltage, the energy balance is broken. Fig.\ref{FigPotential}(c) shows a
potential with a critical voltage. Once, the voltage exceeds the critical
point, the potential has only one minimum without energy barrier as shown in
Fig.\ref{FigPotential}(d). Then, the inner plate moves so that the position $%
u$\ is identical to the positions of the outermost plates, i.e., $%
u=u_{1}=u_{3}=X_{0}$. It corresponds to the FM interaction in the Ising
model.

We derive the critical voltage as in the case of the MEMS memory. By using
an approximation $X_{\text{cap}}\gg u$, the static energy is approximated as%
\begin{equation}
U_{\text{cap}}\simeq \frac{\varepsilon _{0}SV^{2}X_{\text{cap}}}{2\left( X_{%
\text{cap}}^{2}-X_{0}^{2}\right) }\left( 1-\frac{2X_{0}}{X_{\text{cap}%
}^{2}-X_{0}^{2}}u\right) .
\end{equation}%
The stationary solution $u_{0}$\ is obtained by solving Eq.(\ref{Uuu}), or 
\begin{equation}
\frac{\partial ^{2}U_{\text{total}}}{\partial u^{2}}=4\alpha u^{2}+2\alpha
\left( u^{2}-X_{0}^{2}\right) =0.
\end{equation}%
It yields $u_{0}=-X_{0}/\sqrt{3}$. Then, the critical value is obtained from
(\ref{EqE}), or%
\begin{eqnarray}
\left. \frac{\partial U_{\text{total}}}{\partial u}\right\vert _{u_{0}}
&=&\left. 2\alpha u\left( u^{2}-X_{0}^{2}\right) -\frac{2\varepsilon _{0}SX_{%
\text{cap}}X_{0}}{\left( X_{\text{cap}}^{2}-X_{0}^{2}\right) ^{2}}%
V^{2}\right\vert _{u_{0}}  \notag \\
&=&\frac{4\alpha X_{0}^{3}}{3\sqrt{3}}-\frac{2\varepsilon _{0}SX_{\text{cap}%
}X_{0}}{\left( X_{\text{cap}}^{2}-X_{0}^{2}\right) ^{2}}V^{2}<0.
\end{eqnarray}%
It yields $V>V_{\text{cr}}$\ with%
\begin{equation}
V_{\text{cr}}=\sqrt{\frac{2\alpha X_{0}^{2}\left( X_{\text{cap}%
}^{2}-X_{0}^{2}\right) ^{2}}{3\sqrt{3}\varepsilon _{0}SX_{\text{cap}}}}.
\label{MemsCrilV}
\end{equation}%
This is Eq.(\ref{CriticalV}) in the main text. It is necessary to apply a
voltage larger than the critical voltage $V_{\text{cr}}$\ in order to carry
out the MEMS computing.

\section{Derivation of Ising model}

\label{Rep}

The potential energy of the buckled-plate MEMS is given by 
\begin{equation}
U_{\text{total}}=U_{\text{mech}}+U_{\text{cap}},  \label{EneMEMS}
\end{equation}%
where $U_{\text{mech}}$ is given by Eq.(\ref{Pote}) and%
\begin{equation}
U_{\text{cap}}=\sum_{j=1}^{N-1}\frac{C_{\text{para}}(u_{j},u_{j+1})}{2}%
V_{j}^{2}.  \label{EneN}
\end{equation}%
Provided the applied voltage $V_{j}$ is sufficiently smaller than $V_{\text{%
cr}}$, the ground state is given by Eq.(\ref{Mini}), or $s_{j}\equiv
u_{j}/X_{0}=\pm 1$.

We show that the potential energy (\ref{EneMEMS}) may be written in the form
of the Ising model under external field $B_{j}$,%
\begin{equation}
H_{\text{Ising}}=\sum_{j=1}^{N-1}J_{j}s_{j}s_{j+1}+%
\sum_{j=1}^{N}B_{j}s_{j}+E_{0}.  \label{IsingA}
\end{equation}%
We rewrite Eq.(\ref{IsingA}) as%
\begin{equation}
H_{\text{Ising}}=\sum_{j=1}^{N-1}H_{j}\left( s_{j},s_{j+1}\right) +\frac{%
B_{1}}{2}s_{1}+\frac{B_{N}}{2}s_{N},  \label{HHs}
\end{equation}%
with%
\begin{equation}
H_{j}\left( s_{j},s_{j+1}\right) =J_{j}s_{j}s_{j+1}+\frac{B_{j}}{2}s_{j}+%
\frac{B_{j+1}}{2}s_{j+1}+\frac{E_{0}}{N-1}.  \label{Hs}
\end{equation}%
We realize the term (\ref{Hs}) by a system made of two adjacent buckled
plates $j$ and $j+1$.

The energy $H_{j}\left( s_{j},s_{j+1}\right) $ of the Ising Hamiltonian is
given by setting $s_{j}=\pm 1$ in Eq.(\ref{Hs}), while the energy $U_{\text{%
cap}}$ of the buckled-plate MEMS is given by setting $u_{j}=X_{0}s_{j}$ in
Eq.(\ref{EneN}) provided $V/V_{\text{cr}}\ll 1$. By identifying them, we
obtain a formula%
\begin{equation}
H_{j}\left( s_{j},s_{j+1}\right) =\frac{C_{\text{para}}(u_{j},u_{j+1})}{2}%
V_{j}^{2}  \label{EqA}
\end{equation}%
with $u_{j}=X_{0}s_{j}$ under the constant-voltage condition. Note that we
have $U_{\text{mech}}=0$ since the plates exist at the stable position. The
condition (\ref{EqA}) determines the parameters $J_{j}$, $B_{j}$\ and $E_{0}$%
\ in Eq.(\ref{IsingA}).

We write down $H_{j,j+1}\left( s_{j},s_{j+1}\right) $ explicitly,%
\begin{align}
H_{j}\left( 1,1\right) & =J_{j}+B_{j}^{\prime }+B_{j+1}^{\prime
}+E_{0}^{\prime }\equiv \mathcal{E}_{0}, \\
H_{j}\left( 1,-1\right) & =-J_{j}+B_{j}^{\prime }-B_{j+1}^{\prime
}+E_{0}^{\prime }\equiv \mathcal{E}_{+}, \\
H_{j}\left( -1,1\right) & =-J_{j}-B_{j}^{\prime }+B_{j+1}^{\prime
}+E_{0}^{\prime }\equiv \mathcal{E}_{-}, \\
H_{j}\left( -1,-1\right) & =J_{j}-B_{j}^{\prime }-B_{j+1}^{\prime
}+E_{0}^{\prime }\equiv \mathcal{E}_{0},
\end{align}%
where we have defined $B_{j}^{\prime }=B_{j}/2$ and $E_{0}^{\prime
}=E_{0}/\left( N-1\right) $. It follows from Eqs.(\ref{EqA}) and (\ref{Cu})
that%
\begin{align}
\mathcal{E}_{0}& =\frac{\varepsilon _{0}S}{X_{\text{cap}}-X_{0}+X_{0}}\frac{%
V_{j}^{2}}{2}=\frac{\varepsilon _{0}S}{X_{\text{cap}}}\frac{V_{j}^{2}}{2},
\label{EqAA} \\
\mathcal{E}_{\pm }& =\frac{\varepsilon _{0}S}{X_{\text{cap}}\mp X_{0}\mp
X_{0}}\frac{V_{j}^{2}}{2}=\frac{\varepsilon _{0}S}{X_{\text{cap}}\mp 2X_{0}}%
\frac{V_{j}^{2}}{2}.  \label{EqAAA}
\end{align}%
We find $\mathcal{E}_{+}>\mathcal{E}_{0}>\mathcal{E}_{-}$. Hence, the
coefficients in the Ising model are given%
\begin{align}
J_{j}& =\frac{2\mathcal{E}_{0}-\mathcal{E}_{+}-\mathcal{E}_{-}}{4}=-\frac{%
2\varepsilon _{0}SV_{j}^{2}X_{0}^{2}}{X_{\text{cap}}\left( X_{\text{cap}%
}^{2}-4X_{0}^{2}\right) },  \label{EqB} \\
B_{j}& =-B_{j+1}=\frac{\mathcal{E}_{+}-\mathcal{E}_{-}}{4}=\frac{\varepsilon
_{0}SV_{j}^{2}X_{0}}{\left( X_{\text{cap}}^{2}-4X_{0}^{2}\right) },
\label{EqC} \\
E_{0}& =\frac{2\mathcal{E}_{0}+\mathcal{E}_{+}+\mathcal{E}_{-}}{4}=\frac{%
2\varepsilon _{0}SV_{j}^{2}\left( X_{\text{cap}}^{2}-X_{0}^{2}\right) }{X_{%
\text{cap}}\left( X_{\text{cap}}^{2}-4X_{0}^{2}\right) },  \label{EqD}
\end{align}%
in terms of the quantities of the buckled-plate MEMS.

We recall that we apply a constant voltage $V$\ at the odd plates while the
odd plates are grounded in the Ising machine. Then, the voltage is given by $%
V_{2j-1}=-V$ and $V_{2j}=V$. See Fig.\ref{FigBoundary}(a). It results in the
constant voltage $V_{j}^{2}=V^{2}$. The Hamiltonian (\ref{IsingA}) is
reduced to%
\begin{equation}
H_{\text{Ising}}=%
\sum_{j=1}^{N-1}J_{j}s_{j}s_{j+1}+B_{1}s_{1}+B_{N}s_{N}+E_{0},  \label{Hamil}
\end{equation}%
since the $B_{j}s_{j}$ terms cancel except for the end plates. Here, $J_{j}$%
, $B_{1}=-B_{N}$ and $E_{0}$ are given by Eqs.(\ref{EqB}), (\ref{EqC}) and (%
\ref{EqD}), respectively.

\begin{figure}[t]
\centerline{\includegraphics[width=0.48\textwidth]{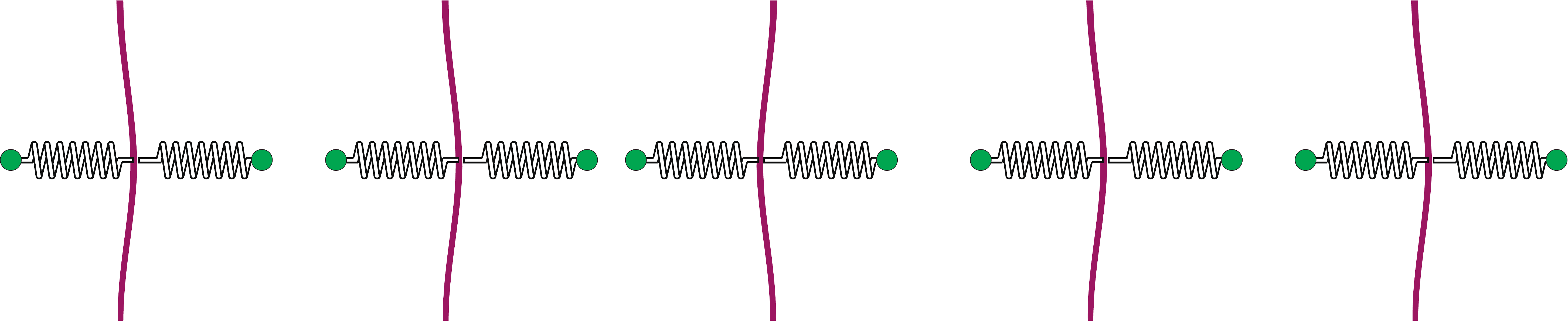}}
\caption{Illustration of plates connected by springs. Each buckled plate is
connected by a spring where the other end is fixed. This structure realizes
the pure Ising model without external field term $B_{j}s_{j}$.}
\label{FigSpringPlate}
\end{figure}

However, $X_{\text{cap}}$ is inhomogeneous in the presence of randomness,
where the $B_{j}s_{j}$ terms do not cancel. In order to control the $%
B_{j}s_{j}$\ term independently, we introduce an additional spring
connecting a plate and a fixed point as in Fig.\ref{FigSpringPlate}, leading
to the potential energy given by%
\begin{equation}
U_{\text{add}}=\frac{\kappa }{2}\sum_{j=1}^{N}\left( u_{j}-X_{j}^{\text{add}%
}\right) ^{2}.
\end{equation}%
The modified Ising Hamiltonian is 
\begin{equation}
H_{\text{Ising}}^{\prime }=H_{\text{Ising}}+\sum_{j=1}^{N}H_{\text{add}%
}\left( s_{j}\right) ,
\end{equation}%
with%
\begin{align}
H_{\text{add}}\left( s_{j}\right) =&\frac{\kappa }{2}\left(
s_{j}X_{0}-X_{j}^{\text{add}}\right) ^{2}  \notag \\
=&\kappa X_{0}X_{j}^{\text{add}}s_{j}+\frac{\kappa }{2}\left(
X_{0}^{2}+\left( X_{j}^{\text{add}}\right) ^{2}\right) .
\end{align}%
It is rewritten in the form%
\begin{equation}
H_{\text{add}}\left( s_{j}\right) =B_{j}^{\text{add}}s_{j}+E_{j}^{\text{add}%
},
\end{equation}%
with%
\begin{equation}
B_{j}^{\text{add}}=\kappa X_{0}X_{j}^{\text{add}},\quad E_{j}^{\text{add}}=%
\frac{\kappa }{2}\left( X_{0}^{2}+\left( X_{j}^{\text{add}}\right)
^{2}\right) .
\end{equation}%
Hence, it is possible to control the $B_{j}s_{j}$ term by using the $B_{j}^{%
\text{add}}s_{j}$ term because $X_{\text{add}}$ can be positive or negative.

In order to cancel the $B_{j}s_{j}$ term, it is enough to set%
\begin{equation}
\kappa X_{j}^{\text{add}}=\frac{\varepsilon _{0}SV_{j}^{2}}{\left( X_{\text{%
cap}}^{2}-4X_{0}^{2}\right) }.
\end{equation}%
where we have used Eq.(\ref{EqC}). We now obtain the pure Ising Hamiltonian, 
\begin{equation}
H_{\text{Ising}}^{\prime }=\sum_{j=1}^{N-1}J_{j}s_{j}s_{j+1}+E_{0},
\label{IsingH}
\end{equation}%
where the $B_{j}s_{j}$ terms are absent. We have shown that the presence of
randomness is remedied by tuning additional springs.

\begin{figure}[t]
\centerline{\includegraphics[width=0.48\textwidth]{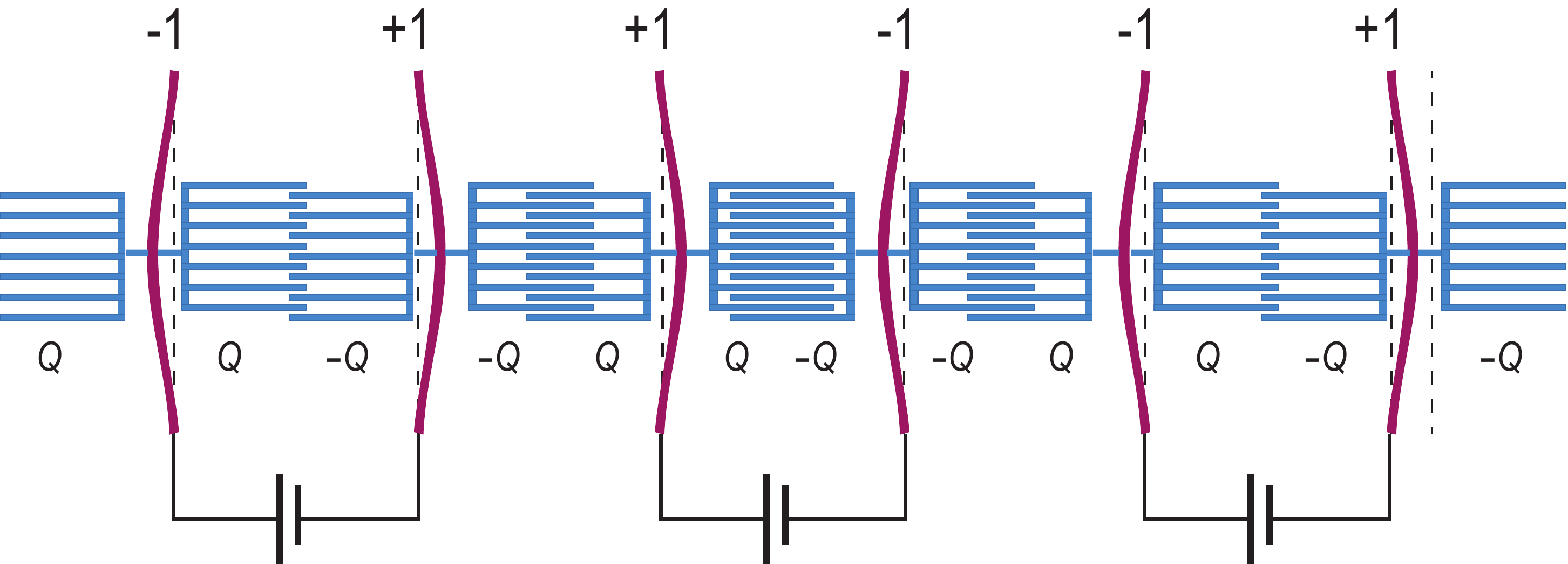}}
\caption{Illustration of a comb-teeth MEMS, where the direction of buckling
is controlled by the charge stored in a comb-teeth MEMS. The charge is
supplied externally to the capacitor. Here a homogeneous charge is assumed.}
\label{FigComb}
\end{figure}

\section{Characteristic frequency}

\label{SecFreq}

We estimate the highest frequency of the dynamics. The highest frequency is
given by the oscillation at the bottom of the potential (\ref{Pote}) at $%
u_{j}=\pm X_{0}$. The potential is expanded around $u_{j}=X_{0}$\ as%
\begin{equation}
U_{\text{mech}}^{\left( j\right) }\equiv \frac{\alpha }{2}\left(
u_{j}^{2}-X_{0}^{2}\right) ^{2}\simeq 2\alpha X_{0}^{2}\left(
u_{j}-X_{0}\right) ^{2}.
\end{equation}%
The equation of motion is given by%
\begin{equation}
m\ddot{u}_{j}+\gamma \dot{u}_{j}=-4\alpha X_{0}^{2}\left( u_{j}-X_{0}\right)
.
\end{equation}%
The characteristic frequency is obtained as%
\begin{equation}
\omega =\sqrt{\frac{4\alpha X_{0}^{2}}{m}}.
\end{equation}

\section{Square Network of MEMS}

\label{SecNetwork}

\begin{figure}[t]
\centerline{\includegraphics[width=0.48\textwidth]{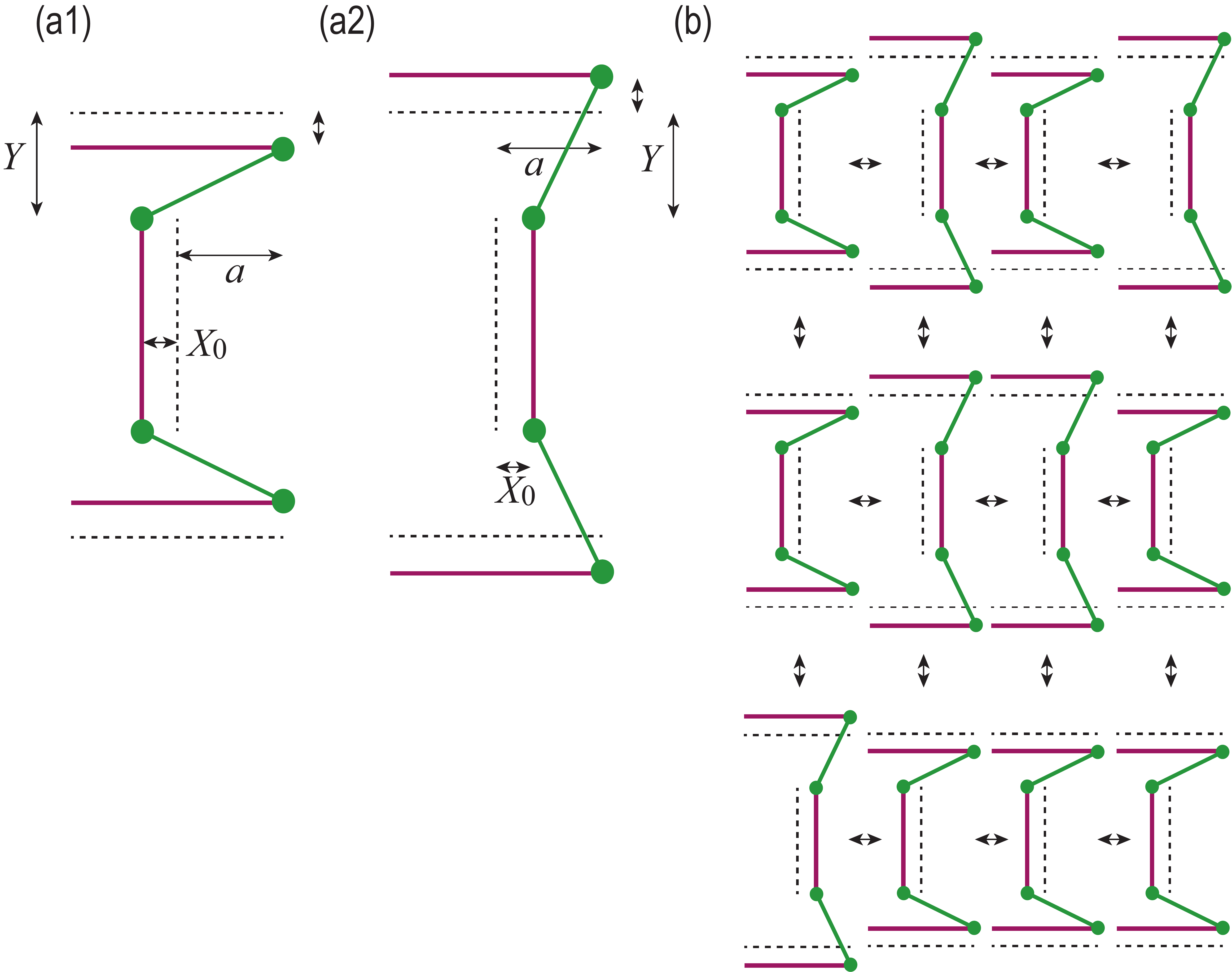}}
\caption{(a1) and (a2) Illustration of a mechanism connecting the horizontal
($x$) and vertical ($y$) motions of MEMS. There are only two configurations,
that is (a1) left-down position and (a2) right-up position. (b) A
two-dimensional network of MEMS. }
\label{FigNetwork}
\end{figure}

A square network is often used in the annealing machine. We show a method to
construct two-dimensional networks of MEMS because there is no frustration
in one dimensional MEMS array. We show a mechanism to transfer the motion
along the $x$ axis to the $y$ axis in Fig.\ref{FigNetwork}(a). The two
plates are connected by a green spring, where the end points are the edges
of plates. The green spring has a natural length $L_{0}$ and rotates freely
at the end points. We assume the\ position of the upper plate is $y=Y$ and
the length of the plate is $2a$. When the position of the lower plate is
left and the position of the upper plate is down as in Fig.\ref{FigNetwork}%
(a1), the length of the green bar is given by 
\begin{equation}
L_{+}=\sqrt{\left( X_{0}+a\right) ^{2}+\left( Y-X_{0}\right) ^{2}}.
\end{equation}%
In the similar way, when the position of the lower plate is right and the
the position of the upper plate is up as in Fig.\ref{FigNetwork}(a2), the
length of the green bar is given by

\begin{equation}
L_{-}=\sqrt{\left( -X_{0}+a\right) ^{2}+\left( Y+X_{0}\right) ^{2}}.
\end{equation}%
By equating $L_{+}=L_{-}$, we find that it enough to set $a=Y$. By using
this condition, the motion of the lower plate along the $x$ axis is
transformed into the $y$ motion of the upper plate.

Next, we consider a network shown in Fig.\ref{FigNetwork}(b). There are
electrostatic interaction both along the $x$ and $y$ axes. It forms a square
network of MEMS.

\section{Comb-teeth MEMS}

\label{SecCombTeeth}

We have so far considered an Ising machine made of the parallel-plate MEMS.
In the current MEMS technology, however, comb-teeth electrostatic actuators
(comb-teeth MEMS) are commonly used instead of parallel plate MEMS. It is
due to the tradeoff of force, maximum stroke vs stability. Parallel plate
actuator can be stably controllable only for $1/3$ of its initial gap $X_{%
\text{cap}}$. As soon as the displacement $u$ exceeds $X_{\text{cap}}/3$,
the plate is continuously pulled until the two plates touch and make
short-circuit. This phenomenon is called the pull-in instability. It is
because the electrostatic force is inversely proportional to the square of
gap: $F(X_{\text{cap}}-u)^{-2}$, whereas the spring counterforces to hold
the plate against pull-in is only proportional to its displacement: $%
F=\kappa u$. On the other hand, the comb-teeth MEMS machine does not suffer
from the pull-in instability because the electrostatic force is constant
regardless of the displacement.

It is unnecessary to use metal for buckled plates. We can use plastic or gum
for the buckled plates, which extends the possibility of the parameter
tuning of the plates. These are merits to overcome the complexity to
manufacture the comb-teeth MEMS.

The structure is shown in Fig.\ref{FigComb}, where many parallel plates form
a comb-teeth like structure. The capacitance of the comb-teeth MEMS is
proportional to the overlapped area, 
\begin{equation}
C_{\text{comb}}(u_{j},u_{j+1})=\frac{\varepsilon _{0}d_{Y}\left( X_{\text{cap%
}}+u_{j}-u_{j+1}\right) }{d_{Z}},
\end{equation}%
where $d_{Z}$ is the distance between the adjacent plates within the
comb-teeth MEMS, and $d_{Y}$\ is the length of the plate of the comb-teeth
MEMS along the $y$ direction.

Here we control charge $Q_{j}$ stored in $j$-th comb-teeth MEMS. The static
energy is given by%
\begin{equation}
U_{\text{comb}}=\sum_{j}\frac{Q_{j}^{2}}{2C_{\text{comb}}(u_{j},u_{j+1})}
\label{Ucomb}
\end{equation}%
in the presence of a charge $Q_{j}$ between the $j$-th and ($j+1$)-th
plates. We consider the configuration shown in Fig.\ref{FigComb}, where the
charge difference is $Q_{j}$ between the adjacent plates. It is possible to
construct an Ising machine based on the comb-teeth MEMS with charge control.

We note that the dynamics is identical between the parallel plates
controlling voltage and the comb-teeth MEMS controlling charge by suitably
changing variables. Namely, there is a kind of duality between the two
systems, as we now show.

\section{Duality}

\label{SecDuality}

We have shown that the Ising machine may be constructed with the use of the
parallel MEMS by controlling voltage. We have also shown that the Ising
machine may be constructed with the use of the comb-teeth MEMS by
controlling charge. We argue that there is a duality relation between these
two systems.

The differences between two types of MEMS are the electrostatic potentials
Eq.(\ref{EneN}) under the constant voltage and Eq.(\ref{Ucomb}) under the
constant charge, i.e., 
\begin{align}
U_{\text{para}}& =\sum_{j}\frac{C_{\text{para}}(u_{j},u_{j+1})}{2}V_{j}^{2},
\\
U_{\text{comb}}& =\sum_{j}\frac{Q_{j}^{2}}{2C_{\text{comb}}(u_{j},u_{j+1})}.
\end{align}%
They are explicitly given by%
\begin{align}
U_{\text{para}}& =\frac{\varepsilon _{0}S}{X_{\text{cap}}+u_{j}-u_{j+1}}%
\frac{V_{j}^{2}}{2}, \\
U_{\text{cap}}& =\frac{d_{Z}}{\varepsilon _{0}d_{Y}\left( X_{\text{cap}%
}+u_{j}-u_{j+1}\right) }\frac{Q_{j}^{2}}{2}.
\end{align}%
These two potentials are transformed into one another by%
\begin{equation}
V_{j}\Leftrightarrow Q_{j},\qquad \varepsilon _{0}S\Leftrightarrow \frac{%
d_{Z}}{\varepsilon _{0}d_{Y}}.  \label{VQ}
\end{equation}%
This is the duality relation between the parallel-plate MEMS and the
comb-teeth MEMS. All numerical results of the parallel-plate MEMS are
applicable to the comb-teeth MEMS by changing parameters according to Eq.(%
\ref{VQ}).

\begin{widetext}
\section{Explicit equations of motions for fully-connected systems}

\red{The equations of motion corresponding to Fig.\ref{FigFull}(a4) are given
by%
\begin{align}
m\ddot{u}_{1}+\gamma \dot{u}_{1}+2\alpha u_{1}\left( u_{1}^{2}-X_{0}\right)
& = & & \frac{\varepsilon _{0}SV_{12}^{2}}{\left( X_{\text{cap}%
}+u_{1}-u_{2}\right) ^{2}}+\frac{\varepsilon _{0}SV_{13}^{2}}{\left( X_{%
\text{cap}}+u_{1}-u_{3}\right) ^{2}}+\frac{\varepsilon _{0}SV_{14}^{2}}{%
\left( X_{\text{cap}}+u_{1}-u_{4}\right) ^{2}}+\frac{\varepsilon
_{0}SV_{15}^{2}}{\left( X_{\text{cap}}+u_{1}-u_{5}\right) ^{2}}, \\
m\ddot{u}_{2}+\gamma \dot{u}_{2}+2\alpha u_{2}\left( u_{2}^{2}-X_{0}\right)
& = & & -\frac{\varepsilon _{0}SV_{12}^{2}}{\left( X_{\text{cap}%
}+u_{1}-u_{2}\right) ^{2}}+\frac{\varepsilon _{0}SV_{23}^{2}}{\left( X_{%
\text{cap}}+u_{2}-u_{3}\right) ^{2}}+\frac{\varepsilon _{0}SV_{24}^{2}}{%
\left( X_{\text{cap}}+u_{2}-u_{4}\right) ^{2}}+\frac{\varepsilon
_{0}SV_{25}^{2}}{\left( X_{\text{cap}}+u_{2}-u_{5}\right) ^{2}}, \\
m\ddot{u}_{3}+\gamma \dot{u}_{3}+2\alpha u_{3}\left( u_{3}^{2}-X_{0}\right)
& = & & -\frac{\varepsilon _{0}SV_{13}^{2}}{\left( X_{\text{cap}%
}+u_{1}-u_{3}\right) ^{2}}-\frac{\varepsilon _{0}SV_{23}^{2}}{\left( X_{%
\text{cap}}+u_{2}-u_{3}\right) ^{2}}+\frac{\varepsilon _{0}SV_{34}^{2}}{%
\left( X_{\text{cap}}+u_{3}-u_{4}\right) ^{2}}+\frac{\varepsilon
_{0}SV_{35}^{2}}{\left( X_{\text{cap}}+u_{3}-u_{5}\right) ^{2}}, \\
m\ddot{u}_{4}+\gamma \dot{u}_{4}+2\alpha u_{4}\left( u_{4}^{2}-X_{0}\right)
& = & & -\frac{\varepsilon _{0}SV_{14}^{2}}{\left( X_{\text{cap}%
}+u_{1}-u_{4}\right) ^{2}}-\frac{\varepsilon _{0}SV_{24}^{2}}{\left( X_{%
\text{cap}}+u_{2}-u_{4}\right) ^{2}}-\frac{\varepsilon _{0}SV_{34}^{2}}{%
\left( X_{\text{cap}}+u_{3}-u_{4}\right) ^{2}}+\frac{\varepsilon
_{0}SV_{45}^{2}}{\left( X_{\text{cap}}+u_{4}-u_{5}\right) ^{2}}, \\
m\ddot{u}_{5}+\gamma \dot{u}_{5}+2\alpha u_{5}\left( u_{5}^{2}-X_{0}\right)
& = & & -\frac{\varepsilon _{0}SV_{15}^{2}}{\left( X_{\text{cap}%
}+u_{1}-u_{5}\right) ^{2}}-\frac{\varepsilon _{0}SV_{25}^{2}}{\left( X_{%
\text{cap}}+u_{2}-u_{5}\right) ^{2}}-\frac{\varepsilon _{0}SV_{35}^{2}}{%
\left( X_{\text{cap}}+u_{3}-u_{5}\right) ^{2}}-\frac{\varepsilon
_{0}SV_{45}^{2}}{\left( X_{\text{cap}}+u_{4}-u_{5}\right) ^{2}}.
\end{align}}

\red{The equations of motion corresponding to Fig.\ref{FigFull}(a5) are given by%
\begin{align}
m\ddot{u}_{1}+\gamma \dot{u}_{1}+2\alpha u_{1}\left( u_{1}^{2}-X_{0}\right)
& = & & -\frac{\varepsilon _{0}SV_{j}^{2}}{\left( X_{\text{cap}%
}+u_{1}-u_{2}\right) ^{2}}+\frac{\varepsilon _{0}SV_{j}^{2}}{\left( X_{\text{%
cap}}+u_{1}-u_{3}\right) ^{2}}+\frac{\varepsilon _{0}SV_{j}^{2}}{\left( X_{%
\text{cap}}+u_{1}-u_{4}\right) ^{2}}+\frac{\varepsilon _{0}SV_{j}^{2}}{%
\left( X_{\text{cap}}+u_{1}-u_{5}\right) ^{2}}, \\
m\ddot{u}_{2}+\gamma \dot{u}_{2}+2\alpha u_{2}\left( u_{2}^{2}-X_{0}\right)
& = & & -\frac{\varepsilon _{0}SV_{j}^{2}}{\left( X_{\text{cap}%
}+u_{2}-u_{1}\right) ^{2}}+\frac{\varepsilon _{0}SV_{j}^{2}}{\left( X_{\text{%
cap}}+u_{2}-u_{3}\right) ^{2}}+\frac{\varepsilon _{0}SV_{j}^{2}}{\left( X_{%
\text{cap}}+u_{2}-u_{4}\right) ^{2}}+\frac{\varepsilon _{0}SV_{j}^{2}}{%
\left( X_{\text{cap}}+u_{2}-u_{5}\right) ^{2}}, \\
m\ddot{u}_{3}+\gamma \dot{u}_{3}+2\alpha u_{3}\left( u_{3}^{2}-X_{0}\right)
& = & & -\frac{\varepsilon _{0}SV_{j}^{2}}{\left( X_{\text{cap}%
}+u_{1}-u_{3}\right) ^{2}}-\frac{\varepsilon _{0}SV_{j}^{2}}{\left( X_{\text{%
cap}}+u_{2}-u_{3}\right) ^{2}}+\frac{\varepsilon _{0}SV_{j}^{2}}{\left( X_{%
\text{cap}}+u_{3}-u_{4}\right) ^{2}}+\frac{\varepsilon _{0}SV_{j}^{2}}{%
\left( X_{\text{cap}}+u_{3}-u_{5}\right) ^{2}}, \\
m\ddot{u}_{4}+\gamma \dot{u}_{4}+2\alpha u_{4}\left( u_{4}^{2}-X_{0}\right)
& = & & -\frac{\varepsilon _{0}SV_{j}^{2}}{\left( X_{\text{cap}%
}+u_{1}-u_{4}\right) ^{2}}-\frac{\varepsilon _{0}SV_{j}^{2}}{\left( X_{\text{%
cap}}+u_{2}-u_{4}\right) ^{2}}-\frac{\varepsilon _{0}SV_{j}^{2}}{\left( X_{%
\text{cap}}+u_{3}-u_{4}\right) ^{2}}+\frac{\varepsilon _{0}SV_{j}^{2}}{%
\left( X_{\text{cap}}+u_{4}-u_{5}\right) ^{2}}, \\
m\ddot{u}_{5}+\gamma \dot{u}_{5}+2\alpha u_{5}\left( u_{5}^{2}-X_{0}\right)
& = & & -\frac{\varepsilon _{0}SV_{j}^{2}}{\left( X_{\text{cap}%
}+u_{1}-u_{5}\right) ^{2}}-\frac{\varepsilon _{0}SV_{j}^{2}}{\left( X_{\text{%
cap}}+u_{2}-u_{5}\right) ^{2}}-\frac{\varepsilon _{0}SV_{j}^{2}}{\left( X_{%
\text{cap}}+u_{3}-u_{5}\right) ^{2}}-\frac{\varepsilon _{0}SV_{j}^{2}}{%
\left( X_{\text{cap}}+u_{4}-u_{5}\right) ^{2}}.
\end{align}}

\red{The equations of motion corresponding to Fig.\ref{FigFull}(a6) are given by%
\begin{align}
m\ddot{u}_{1}+\gamma \dot{u}_{1}+2\alpha u_{1}\left( u_{1}^{2}-X_{0}\right)
& = & & -\frac{\varepsilon _{0}SV_{j}^{2}}{\left( X_{\text{cap}%
}+u_{1}-u_{2}\right) ^{2}}+\frac{\varepsilon _{0}SV_{j}^{2}}{\left( X_{\text{%
cap}}+u_{1}-u_{3}\right) ^{2}}+\frac{\varepsilon _{0}SV_{j}^{2}}{\left( X_{%
\text{cap}}+u_{1}-u_{4}\right) ^{2}}+\frac{\varepsilon _{0}SV_{j}^{2}}{%
\left( X_{\text{cap}}+u_{1}-u_{5}\right) ^{2}}, \\
m\ddot{u}_{2}+\gamma \dot{u}_{2}+2\alpha u_{2}\left( u_{2}^{2}-X_{0}\right)
& = & & -\frac{\varepsilon _{0}SV_{j}^{2}}{\left( X_{\text{cap}%
}+u_{2}-u_{1}\right) ^{2}}+\frac{\varepsilon _{0}SV_{j}^{2}}{\left( X_{\text{%
cap}}+u_{2}-u_{3}\right) ^{2}}+\frac{\varepsilon _{0}SV_{j}^{2}}{\left( X_{%
\text{cap}}+u_{2}-u_{4}\right) ^{2}}+\frac{\varepsilon _{0}SV_{j}^{2}}{%
\left( X_{\text{cap}}+u_{2}-u_{5}\right) ^{2}}, \\
m\ddot{u}_{3}+\gamma \dot{u}_{3}+2\alpha u_{3}\left( u_{3}^{2}-X_{0}\right)
& = & & -\frac{\varepsilon _{0}SV_{j}^{2}}{\left( X_{\text{cap}%
}+u_{1}-u_{3}\right) ^{2}}-\frac{\varepsilon _{0}SV_{j}^{2}}{\left( X_{\text{%
cap}}+u_{2}-u_{3}\right) ^{2}}+\frac{\varepsilon _{0}SV_{j}^{2}}{\left( X_{%
\text{cap}}+u_{3}-u_{4}\right) ^{2}}+\frac{\varepsilon _{0}SV_{j}^{2}}{%
\left( X_{\text{cap}}+u_{3}-u_{5}\right) ^{2}}, \\
m\ddot{u}_{4}+\gamma \dot{u}_{4}+2\alpha u_{4}\left( u_{4}^{2}-X_{0}\right)
& = & & -\frac{\varepsilon _{0}SV_{j}^{2}}{\left( X_{\text{cap}%
}+u_{1}-u_{4}\right) ^{2}}-\frac{\varepsilon _{0}SV_{j}^{2}}{\left( X_{\text{%
cap}}+u_{2}-u_{4}\right) ^{2}}-\frac{\varepsilon _{0}SV_{j}^{2}}{\left( X_{%
\text{cap}}+u_{3}-u_{4}\right) ^{2}}+\frac{\varepsilon _{0}SV_{j}^{2}}{%
\left( X_{\text{cap}}+u_{4}-u_{5}\right) ^{2}}, \\
m\ddot{u}_{5}+\gamma \dot{u}_{5}+2\alpha u_{5}\left( u_{5}^{2}-X_{0}\right)
& = & & -\frac{\varepsilon _{0}SV_{j}^{2}}{\left( X_{\text{cap}%
}+u_{1}-u_{5}\right) ^{2}}-\frac{\varepsilon _{0}SV_{j}^{2}}{\left( X_{\text{%
cap}}+u_{2}-u_{5}\right) ^{2}}-\frac{\varepsilon _{0}SV_{j}^{2}}{\left( X_{%
\text{cap}}+u_{3}-u_{5}\right) ^{2}}-\frac{\varepsilon _{0}SV_{j}^{2}}{%
\left( X_{\text{cap}}+u_{5}-u_{4}\right) ^{2}}.
\end{align}}%
\end{widetext}

\end{document}